\documentclass[twocolumn]{aastex701}

\usepackage{amsmath} 
\usepackage[version=4]{mhchem}
\usepackage{siunitx}
\begin{document}

\title{The Impact of Irradiation on the Radius and Thermal Evolution of Transiting Brown Dwarfs
}

\author[0000-0003-1622-1302]{Sagnick Mukherjee}\email[show]{samukher@ucsc.edu}
\altaffiliation{51 Pegasi b Fellow} 
\affiliation{School of Earth and Space Exploration, Arizona State University, Tempe, AZ, USA \\ }
\affiliation{Department of Astronomy and Astrophysics, University of California, Santa Cruz, CA 95064, USA \\ }

\author[0000-0002-9843-4354]{Jonathan J. Fortney}
\affiliation{Department of Astronomy and Astrophysics, University of California, Santa Cruz, CA 95064, USA \\ }\email{jfortney@ucsc.edu}

\author[0000-0001-6416-1274, sname=Carmichael, gname=Theron]{Theron W. Carmichael}
\altaffiliation{NSF Ascend Fellow}
\affiliation{Institute for Astronomy, University of Hawai‘i, 2680 Woodlawn Drive, Honolulu, HI 96822, USA}
\email{tcarmich@hawaii.edu}

\author[0009-0005-3386-6091]{C.  Evan Davis}
\affiliation{Department of Astronomy and Astrophysics, University of California, Santa Cruz, CA 95064, USA \\ }\email{a@b.com}

\author[0000-0002-5113-8558]{Daniel P. Thorngren}
\affiliation{William H. Miller III Department of Physics \& Astronomy, Johns Hopkins University, 3400 N Charles St, Baltimore, MD 21218, USA} \email{dpthorngren@gmail.com}


\begin{abstract}
Masses and radii of transiting brown dwarfs can be measured directly in contrast to isolated field brown dwarfs, whose mass and radius inferences are model dependent. Therefore, transiting brown dwarfs are a testbed for the interior and evolutionary models of brown dwarfs and giant exoplanets. We have developed atmospheric and evolutionary models for this emerging population. We show that intense stellar irradiation can cause a large enhancement in the radius of transiting brown dwarfs at all masses, especially if the incident flux exceeds $log_{10}(F/cgs)\ge$9 ($T_{\rm eq}\ge 1450$ K). Stellar irradiation can significantly alter rates of nuclear burning in irradiated brown dwarfs, making the Deuterium-burning and Hydrogen-burning minimum masses strong functions of incident stellar flux. We show that the D-burning and H-burning minimum masses can decrease by 16\% and 13\%, respectively, between isolated and strongly irradiated brown dwarfs ( $log_{10}(F/cgs)\ge$10 ($T_{\rm eq}\ge 2570$ K)). This shows that stellar irradiation has a larger impact on the planet--brown dwarf--star mass boundaries than metallicity or clouds. We show that metal cores or migration affect their evolution to a much lesser extent, whereas low mass highly irradiated old sources can help us test the physics of hot Jupiter radius anomaly. We fit the observed radii of 46 transiting brown dwarfs and show that our irradiated evolutionary models fit their radii better than models that ignore the host star, especially for highly irradiated objects. However, the measured radii of 10 objects are still inconsistent at $>3\sigma$ level, indicating residual gaps in our irradiated evolutionary model.

\end{abstract}



\section{Introduction} 

Understanding the temporal evolution and interior structure of gas giant exoplanets and brown dwarfs is key to interpreting their observed properties such as their radii, luminosities, temperatures, etc., and for probing their formation mechanisms \citep[e.g.,][]{fortney2007planetary,burrows97,saumonmarley08,thorngren16,diamondback,Philips20,miguel23,lopez14}. Significant theoretical and observational efforts have been made to study the evolution and interiors of giant exoplanets and brown dwarfs over the last 30 years. Such efforts have met several challenges on both the theoretical and observational fronts. Key challenges on the theoretical front include uncertainties on the equation of state at high pressure, the amount and spatial distribution of ``metals" in the interior, and the atmospheric properties that control the cooling rate of substellar objects \citep[e.g.,][]{miguel23,fortney10,diamondback,Philips20}. Brown dwarfs have been used as a sample to observationally test these theoretical evolutionary models. However, our inability to directly measure the masses and radii of free-floating field brown dwarfs has somewhat limited this approach. Photometric and spectroscopic observations of field brown dwarfs have traditionally been used to measure their luminosity \citep[e.g.,][]{beiler24,tinney14,dupuy13}. The luminosity measurement along with age constraints has been compared with theoretical evolutionary models to constrain the masses of field brown dwarfs. However, such measurements do not adequately test these theoretical evolutionary models as they are significantly dependent on them. Transiting brown dwarfs, on the other hand, are a sample for which independent, direct, and precise mass and radius measurements are possible. Therefore, they form an ideal sample to test the validity of inputs and assumptions in the evolutionary and interior models of field brown dwarfs and gas giant exoplanets.

Studies have found that short-period brown dwarf companions to main-sequence stars are rare \citep[e.g.,][]{grether06,triaud17}. This rarity has often been called the ``brown dwarf desert". However, missions dedicated to finding transiting systems like {\it TESS} and {\it Kepler} along with radial velocity campaigns have led to a significant growth in the sample of known transiting brown dwarfs to about $\sim$60 objects \citep{Barkaoui25,Carmichael22,Henderson24}. Parallax measurements from {\it Gaia} have led to better constraints on the host star radii of several of these objects, which has also led to higher precision on their companion radius \citep[e.g.,][]{Carmichael22}. The sample of brown dwarfs transiting white dwarf hosts has also been growing over the past few years \citep[e.g.,][]{Casewell2024,Lew22,lothringer20,Casewell20,Casewell20b,Buzard22,Sainsbury-Martinez2021}. These objects are particularly interesting because of the much smaller radii of their host stars than main-sequence hosts, which makes their atmospheric characterization relatively easier. Therefore, the sample of transiting brown dwarfs, both around main-sequence and white dwarf hosts, is now well-suited to test the assumptions and inputs within the evolutionary and interior models of substellar objects \citep[e.g.,][]{beatty14,beatty18,beatty19}, complementing similar work for transiting giant planets \citep[e.g.,][]{Chachan2025,acuna25}.

While several evolutionary models of field brown dwarfs and directly imaged exoplanets already exist \citep[e.g.,][]{Philips20,marley21,diamondback,baraffe15}, evolutionary models for strongly irradiated (transiting) brown dwarfs are largely absent in the literature with some exceptions like \citet{burrows11}. \citet{burrows11} do not include  stellar irradiation effects within their evolutionary models. They instead highlight the role of atmospheric metallicity and cloud opacity on the radius of these objects. Their exclusion of stellar irradiation effects on the evolution of transiting brown dwarfs was based on the findings of \citet{burrows07}, where the effect of the incident stellar flux on the radius of giant planets was found to be diminishing with increasing object mass at an age of $\sim$2.5 Gyrs. However, a relatively narrow incident flux range and age range was considered in these models.

\begin{figure}
    \centering
    \includegraphics[width=1\linewidth]{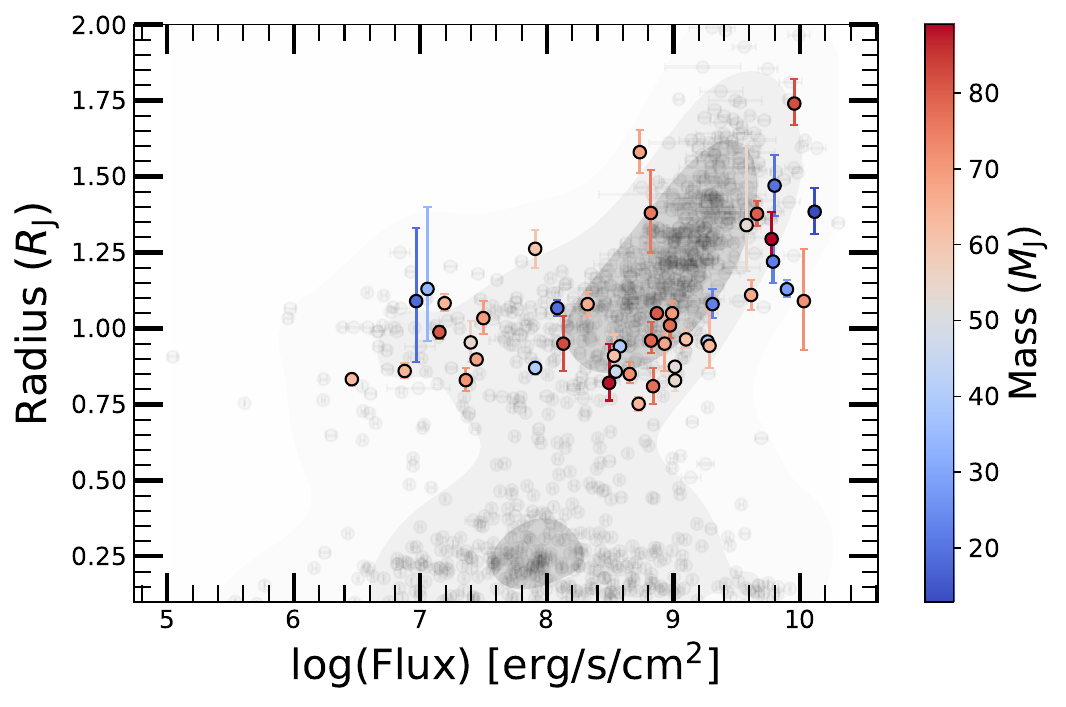}
    \includegraphics[width=1\linewidth]{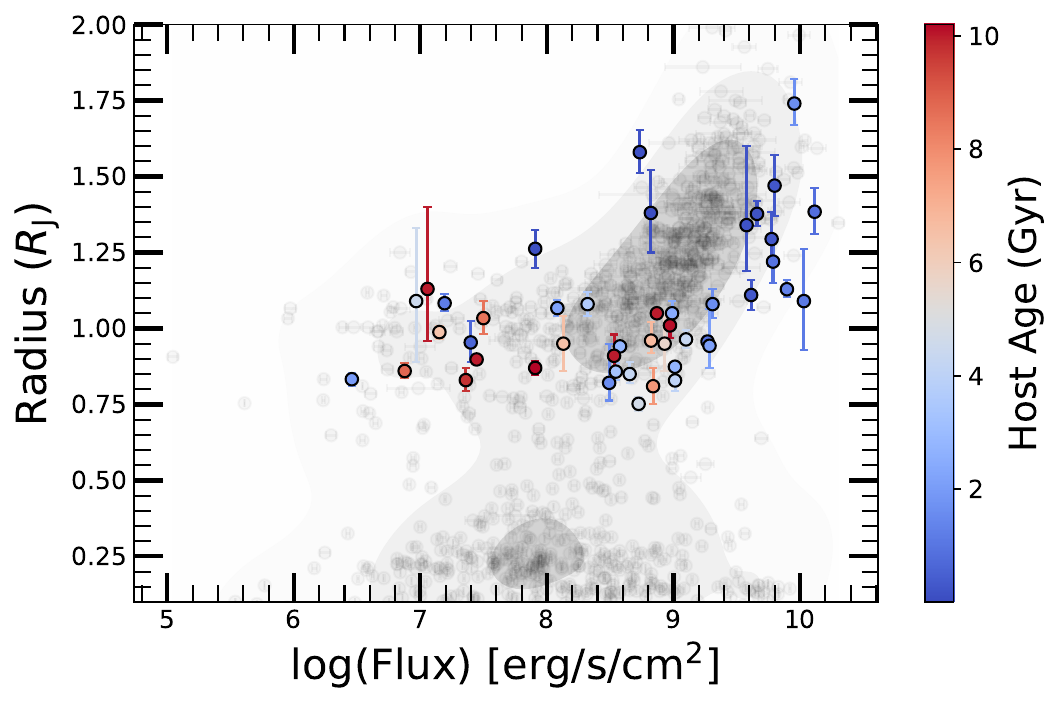}
    \caption{Top panel shows the measured radii of the sample of transiting brown dwarfs as a function of the incident stellar flux on them. The colors of each point corresponds to their mass in Jupiter mass. Bottom panel shows the same quantities in the two axes but the color now corresponds to the median age of the host star of these objects. Objects with masses between 12.9 and 89 $M_{\rm J}$ have been plotted here with colored points whereas the gray points and contours show the transiting exoplanet population with M$<$12.9$ M_{\rm J}$.}
    \label{fig:sample}
\end{figure}

Since then, the sample of known transiting brown dwarf systems has grown considerably, leading to a large range of ages, masses, and incident fluxes within the sample. Figure \ref{fig:sample} shows how the current sample of transiting brown dwarfs (12.9-89 $M_{\rm J}$) show large variations within the radius--incident stellar flux--mass--age parameter space. The incident stellar flux varies by $\sim 4$ orders of magnitude within the sample. Despite this, studies have been forced to use evolutionary models that ignore the effects of host stars to interpret mass and radius measurements of highly irradiated transiting brown dwarfs \citep[e.g.,][]{Benni21,siverd12,Carmichael22, henderson2024, vowell2025,beatty18}. Here, we fill this gap by developing a fully self-consistent and coupled atmospheric and evolutionary model for transiting brown dwarfs using the 1D radiative-convective \texttt{PICASO} atmosphere model and an interior structure and evolutionary model used in the \texttt{SONORA} series \citep{diamondback,evan2025}. We investigate the extent to which various mechanisms affect the evolution of transiting brown dwarfs and their implications for their population. We also apply our models to the measured properties of the latest sample of transiting brown dwarfs to investigate how they match the measurements and identify potential areas where these models need further development.

 In addition to the evolutionary problem, the mass boundaries separating brown dwarfs from planets and brown dwarfs from stars has been debated in the literature over the years \citep[e.g.,][]{burrows97,chabrier2005,spiegel11,kumar63,bodenheimer13,marley21,diamondback,chabrier23,evan2025}. The boundary separating brown dwarfs and planets has been proposed to be based on $^2$H-burning, whereas the boundary between brown dwarfs and stars is based on H-burning. The physical idea behind the planet-brown dwarf boundary is that objects that are massive enough to significantly burn $^2$H in their lifetimes are defined as brown dwarfs. On the other hand, objects that are massive enough such that they can maintain a nearly constant luminosity from H-burning to sufficiently halt their radius contraction are defined as stars. It is now quite clear that these boundaries can depend on the physical state of both the atmospheres and the interiors of substellar objects \citep[e.g.,][]{spiegel11,diamondback,molliere12,bodenheimer13,chabrier23}. Clouds, metal enrichment of atmospheres, equation of state of interiors, and presence of cores have been found to move these boundaries on both ends. However, the mass boundaries between planets and brown dwarfs and brown dwarfs and stars for the class of highly irradiated transiting brown dwarfs have not been well explored. With our evolutionary models, we also investigate these boundaries in detail.

We describe our model in \S\ref{sec:modeling} and present our results in \S\ref{sec:results}. We discuss the areas of future development and applicability of our models in \S\ref{sec:discussion}. Our findings are summarized in \S\ref{sec:conclusions}.

\section{Modeling}\label{sec:modeling}
We divide our transiting brown dwarf evolution model into two components-- 1) interior modeling and 2) atmospheric modeling. We describe each of these components below.

\subsection{Interior \& Evolution}
The evolution and interior structure of substellar objects can be modeled by solving the equations of hydrostatic equilibrium, mass continuity, and energy continuity. These are

\begin{align}
\dfrac{dP}{dm}= -\dfrac{Gm}{4{\pi}r^4} \\
\dfrac{dr}{dm}=\dfrac{1}{4{\pi}{r^2}\rho} \\
\left( L -\int_0^M{\epsilon_{nuc}dm}\right)dt= -\int_0^M{TdSdm}
\end{align}
, respectively. Here, $P$ is the pressure, $m$ is the mass contained within radius $r$, $\rho$ is the density, $L$ is the luminosity, $\epsilon_{nuc}$ is the rate of nuclear energy generation per unit mass, $T$ is the temperature, and $S$ is the specific entropy. We use the numerical model developed and described in \citet{diamondback} \& \citet{thorngren16} to solve these three equations for transiting brown dwarfs. We use the equation of state from \citet{cms19} as input in our evolutionary models. The mass fractions for H, $^2$H, and He in our models are 0.7112, 2.88$\times$10$^{-5}$, and 0.2735, respectively. To compute the nuclear luminosity rate ($\epsilon_{nic}$), we include two reactions following \citet{saumonmarley08},
\begin{align}
    p + p {\rightarrow} d + e^+ + \nu_e \\
    p + d {\rightarrow} ^3He + \gamma
\end{align}
We use the same nuclear reaction cross-sections and screening factors as \citet{saumonmarley08} \& \citet{diamondback} to compute the nuclear luminosity from these reactions.


\subsection{Atmospheres}

\begin{figure*}
    \centering
    \includegraphics[width=1\linewidth]{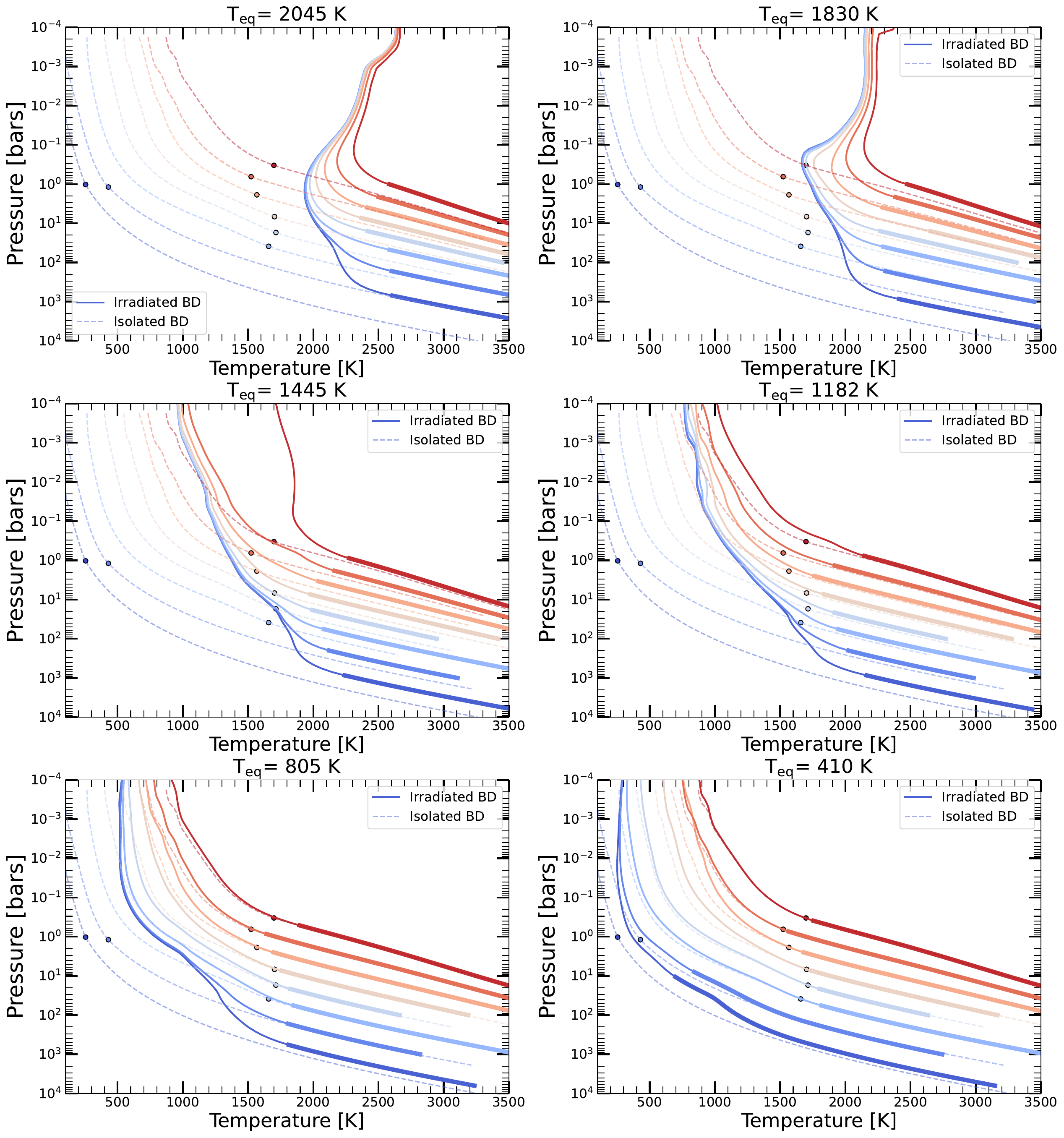}
    \caption{Comparison of $T(P)$ profiles between irradiated brown dwarfs and isolated brown dwarfs. The $T(P)$ profiles of the irradiated brown dwarfs and isolated brown dwarfs are shown with solid and dashed lines, respectively. The deepest convective regions for the irradiated brown dwarfs are shown with thick solid lines and the deepest radiative-convective boundaries for the isolated brown dwarfs are shown with circular points. Each panel corresponds to a different incident stellar flux level. Thermal profiles for $T_{\rm int}$= 300, 500, 800, 1100, 1400, 1700, 2000, and 2300 K are shown in each panel. All $T(P)$ profiles shown here have $log(g)$=5.0. Only the deepest radiative--convective boundaries are denoted here.}
    \label{fig:TP-profile}
\end{figure*}

\begin{figure*}
    \centering
    \includegraphics[width=1\linewidth]{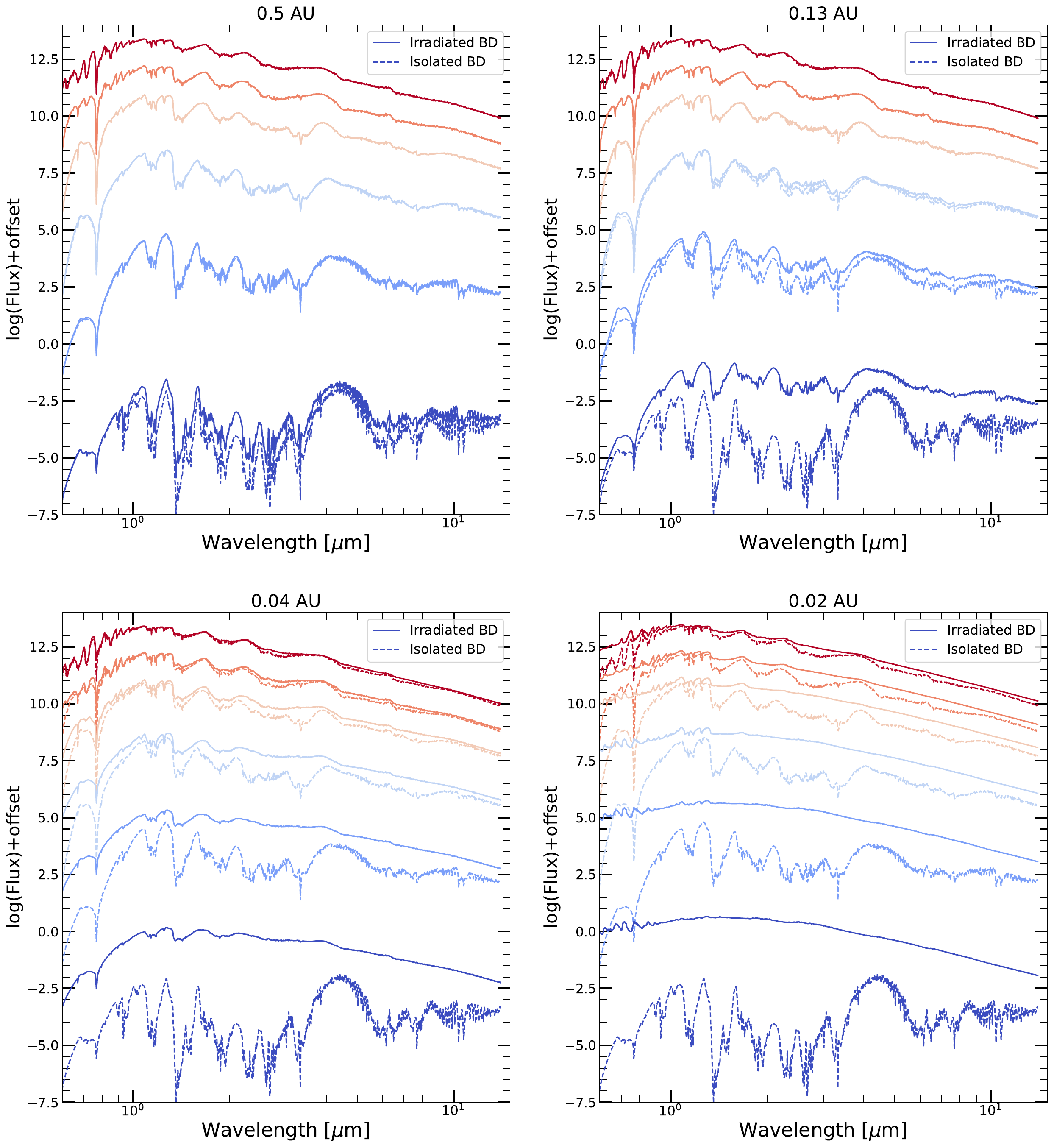}
    \caption{Comparison of emission spectra of irradiated brown dwarfs, to spectra of isolated brown dwarfs from the Sonora bobcat model grid \citep{marley21}. The four panels show model spectra for brown dwarfs located at 0.5, 0.13, 0.04, and 0.02 AU from the host star. The solid lines show the irradiated brown dwarf spectra at $T_{\rm int}$ values between 2400 K and 400 K with a 400 K $T_{\rm int}$ interval in each panel. Solid lines show the spectra of irradiated brown dwarfs whereas the dashed lines show the spectra for an isolated brown dwarf which has the same $T_{\rm eff}$ as the $T_{\rm int}$ of the irradiated brown dwarf model. All models shown here are for $log(g)=5$ and assume thermochemical equilibrium.}
    \label{fig:specta}
\end{figure*}

We use the \texttt{PICASO\footnote{\href{https://natashabatalha.github.io/picaso/}{https://natashabatalha.github.io/picaso/}}} 1D radiative-convective climate model \citep{Mukherjee22,batalha19} to generate model atmospheres across a wide parameter space applicable to transiting brown dwarfs. We assume thermochemical equilibrium and cloud-free conditions throughout the atmosphere to generate these models. Solar composition is assumed with C/O=0.458 \citep{lodders09}. We include gaseous opacities for \ce{CO},  \ce{CH4}, \ce{H2O}, \ce{NH3}, \ce{CO2}, \ce{N2}, \ce{HCN}, \ce{H2}, \ce{C2H2}, \ce{C2H4}, \ce{C2H6}, \ce{Na}, \ce{K}, \ce{PH3}, \ce{TiO}, \ce{VO}, \ce{FeH}, and H$_2$ collision-induced opacity in our climate models and use the resort-rebin technique to mix their correlated-k opacities on-the-fly \citep{amundsen17,Mukherjee22a}. \citet{diamondback} includes a list of the sources of these opacities.

We divide each model atmosphere into 90 plane-parallel layers and assume a Sun-like host star. We compute model atmospheres at 0.02, 0.025, 0.04, 0.06, 0.13, and 0.5 AU from the host star, which correspond to incident fluxes of $log(F/erg.s^{-1}.cm^{-2})$= $9.59$, $9.40$, $8.99$, $8.64$, $7.97$, and $6.80$. We compute model atmospheres for nine $log(g)$ values between 3.5 and 5.5, with an interval of 0.25 in $log(g)$. We include 23 values of $T_{\rm int}$, a parameteriztion of the intrinsic flux from the interior, $\sigma T_{\rm int}^4$ within our atmosphere grid between 200 K and 2400 K. We assume a full redistribution of the incident stellar flux between the day and night sides of the irradiated brown dwarfs, which corresponds to \texttt{rfacv}=0.5 in \texttt{PICASO}. 

Figure \ref{fig:TP-profile} compares the computed $T(P)$ profile for transiting brown dwarfs with $T(P)$ profiles of isolated brown dwarfs from \citet{Mukherjee24}. Each panel shows $T(P)$ profiles computed for different incident flux levels quoted above. The top two panels show large inversions in the irradiated $T(P)$ profiles, which is due to absorption of incident stellar flux by gaseous \ce{TiO} and \ce{VO} \citep{fortney08,hubeny03}. Comparison between the $T(P)$ profiles for high stellar irradiation (top two panels) shows that the stellar irradiation causes the irradiated atmospheres to have significantly hotter adiabats relative to isolated brown dwarfs for a given $T_{\rm int}$. The radiative-convective boundaries of the irradiated brown dwarfs also appear at significantly higher pressures than their non-irradiated counterparts. The extent to which the incident stellar irradiation alters the thermal structure and deep adiabat of irradiated brown dwarfs can be understood with the energy balance equation \citep[e.g.,][]{fortney20},

\begin{align}\label{eq:heat}
    {\sigma}T_{\rm eff}^4 = {\sigma}T_{\rm eq}^4 + {\sigma}T_{\rm int}^4
\end{align}

where the left-hand side term denotes the total flux emitted by the brown dwarf to space, the $T_{\rm eq}$ term denotes the incident stellar flux, and the $T_{\rm int}$ term denotes the heat flux from the object's interior. If the incident stellar flux is significantly greater than the interior heat flux, it significantly alters its thermal and adiabatic structure relative to an isolated brown dwarf model. This is the case for almost all the models shown in the top left panel of Figure \ref{fig:TP-profile}. However, if the interior heat flux is the dominant term in Equation \ref{eq:heat}, then the $T(P)$ profile, the radiative-convective boundary, and the adiabatic structure of the deep atmosphere will be similar between the irradiated and isolated brown dwarf atmospheres. This is the case for the high $T_{\rm int}$ models shown in the bottom two panels in Figure \ref{fig:TP-profile}.

We compare the emission spectra from the irradiated brown dwarf atmospheres models with those for isolated brown dwarfs in Figure \ref{fig:specta}. The four panels show irradiated brown dwarf spectra at four star-brown dwarf separations -- 0.5, 0.13, 0.04, and 0.02 AU. The solid lines in each panel show the irradiated brown dwarf spectra for $T_{\rm int}$ values between 2400 K and 400 K, with a 400 K interval in $T_{\rm int}$. The dashed lines show spectra for isolated brown dwarf atmospheres from the Sonora bobcat model grid \citep{marley21}, where isolated models that have the same $T_{\rm eff}$ values as the $T_{\rm int}$ in the irradiated models are shown with the same color. Figure \ref{fig:specta} shows that the irradiated brown dwarfs show shallower gas absorption features compared to isolated brown dwarf models, which can be attributed to the more isothermal nature of their $T(P)$ profiles in comparison to the isolated brown dwarf models.

The difference in spectra between the two cases is also a strong function of how the $T_{\rm eq}$ of an irradiated brown dwarf model atmosphere compares with $T_{\rm int}$. If $T_{\rm int}\ge{T_{\rm eq}}$, the atmosphere is dominated by the interior heat flux of the object and as a result the spectra is very similar to that of an isolated brown dwarf model that has a $T_{\rm eff}$ similar to the $T_{\rm int}$ of the irradiated brown dwarf. Young transiting brown dwarfs that are well-separated from their host stars can fall in this category.  Among the plotted model spectra in the top left and right panels of Figure \ref{fig:specta}, this is the case for $T_{\rm int}\ge800$ K and $T_{\rm int}\ge1600$ K, respectively. However, if the $T_{\rm eq}$ becomes comparable or greater than the $T_{\rm int}$, then the spectra of a irradiated brown dwarf can be significantly different than an isolated brown dwarf with the same $T_{\rm eff}$ as the $T_{\rm int}$ of the irradiated brown dwarf. Old highly irradiated transiting brown dwarfs can fall in this category. Almost all the plotted models in the bottom two panels of Figure \ref{fig:specta} fall in this category. Moreover, the very strongly irradiated objects with low $T_{\rm int}$ (400 K case in the bottom right panel) show emission lines in their spectra due to strong thermal inversions. 

A key input from the model atmospheres for the interior and evolution model is the emergent heat flux (${\sigma}T_{\rm int}^4$) from the atmosphere for a given gravity and interior structure.  For any atmosphere, we integrate the model down a depth where the atmosphere becomes convective and stays convective, which can be from $<1$ bar to over 100 bar.  This deep convective atmosphere lies on an isentrope, and we can compute the 10-bar temperature of this isentrope, called $T_{10}$.  This serves as the upper boundary condition for the evolution calculation, as the $T_{10}$ is tabulated as a function of $T_{\rm int}$, incident stellar flux $log(F)$, and $log(g)$ \citep[e.g.,][]{saumonmarley08,marley21,diamondback}. This relation from the atmosphere grid is inverted in our interior and evolution model to obtain the emitted heat flux to space from the interior ($T_{\rm int}$) for a given temperature at 10 bars computed from the interior model with a given $log(F)$, and $log(g)$. We present the key results from the evolutionary tracks for transiting brown dwarfs computed using these model atmospheres and interior model in \S\ref{sec:results}.

\section{Results}\label{sec:results}

We explore how various physical factors influence the evolution of transiting brown dwarfs including stellar irradiation, presence of metal cores, and orbital migration. We also explore how our evolutionary models compare with the measured mass, radius, and age of the latest sample of 46 transiting brown dwarfs. 

\subsection{Effects of Stellar Irradiation}

\begin{figure*}
    \centering
    \includegraphics[width=1.0\linewidth]{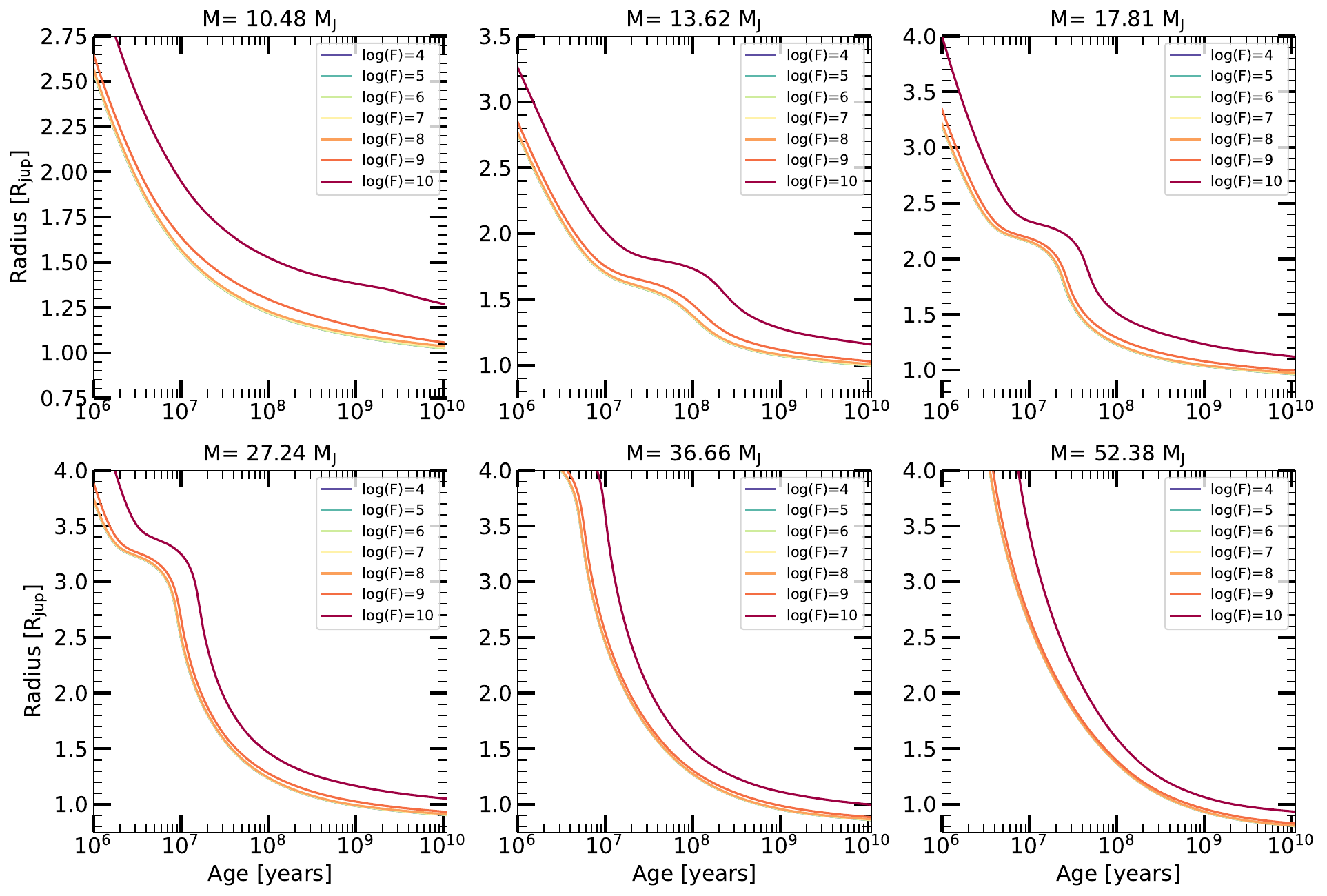}
    \caption{Each panel shows the effect of incident stellar radiation on the radius evolution of irradiated brown dwarfs with varying mass. Radius evolution for 10.48, 13.62, 17.81, 27.24, 36.66, and 52.38 $M_{\rm J}$ are shown in the six panels. The different colored lines in each panel show the evolution due to different incident stellar flux on the object with $log(F)$ varying from 4 to 10, where $F$ is in $erg.s^{-1}.cm^{-2}$. Note that the temporary halting of contraction for M$\ge13.6 M_{\rm J}$ objects is due to the onset of deuterium burning. The radius contraction continues once most of the available deuterium is exhausted.}
    \label{fig:stellar_w_radius}
\end{figure*}

\begin{figure}
    \centering
    \includegraphics[width=1.0\linewidth]{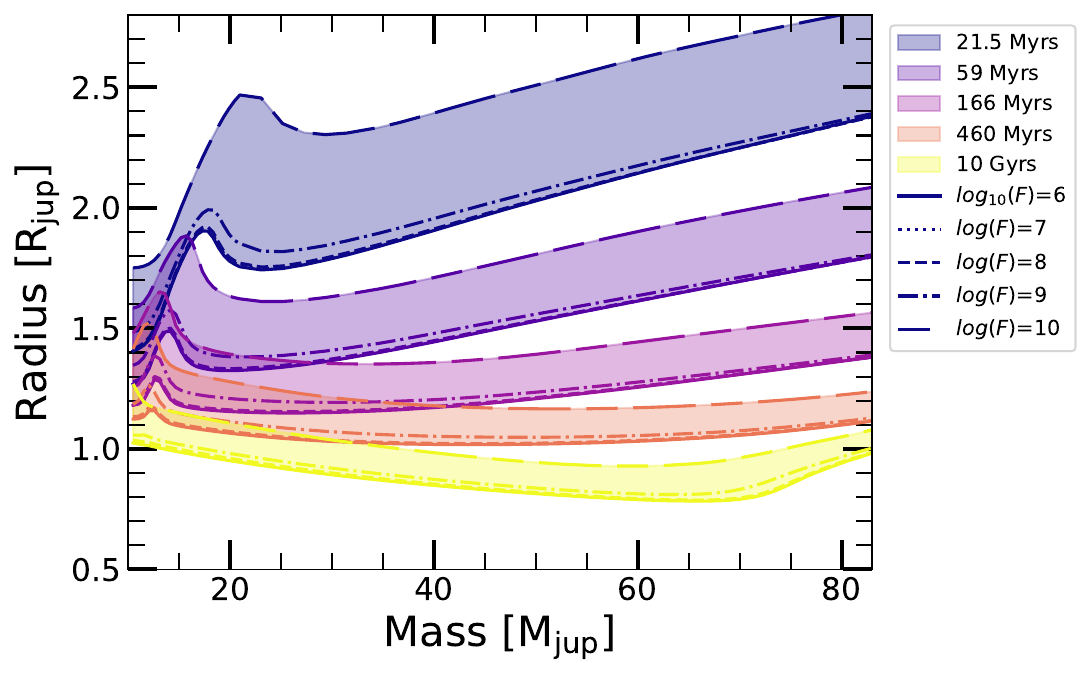}
    \caption{Effect of incident stellar flux on the mass-radius relation of irradiated brown dwarfs at various ages is shown. Each colored shaded region shows the variation in the mass-radius relation due to varying incident stellar flux from $log(F)=6$ to $log(F)=10$. The mass-radius relation for ages 21.5 Myrs, 59 Myrs, 166 Myrs, 480 Myrs, and 10 Gyrs is shown. These ages were chosen to encompass a wide range of ages but still be distinctly visible on the figure.}
    \label{fig:mass_radius_stellar}
\end{figure}

\begin{figure*}
    \centering
    \includegraphics[width=1.0\linewidth]{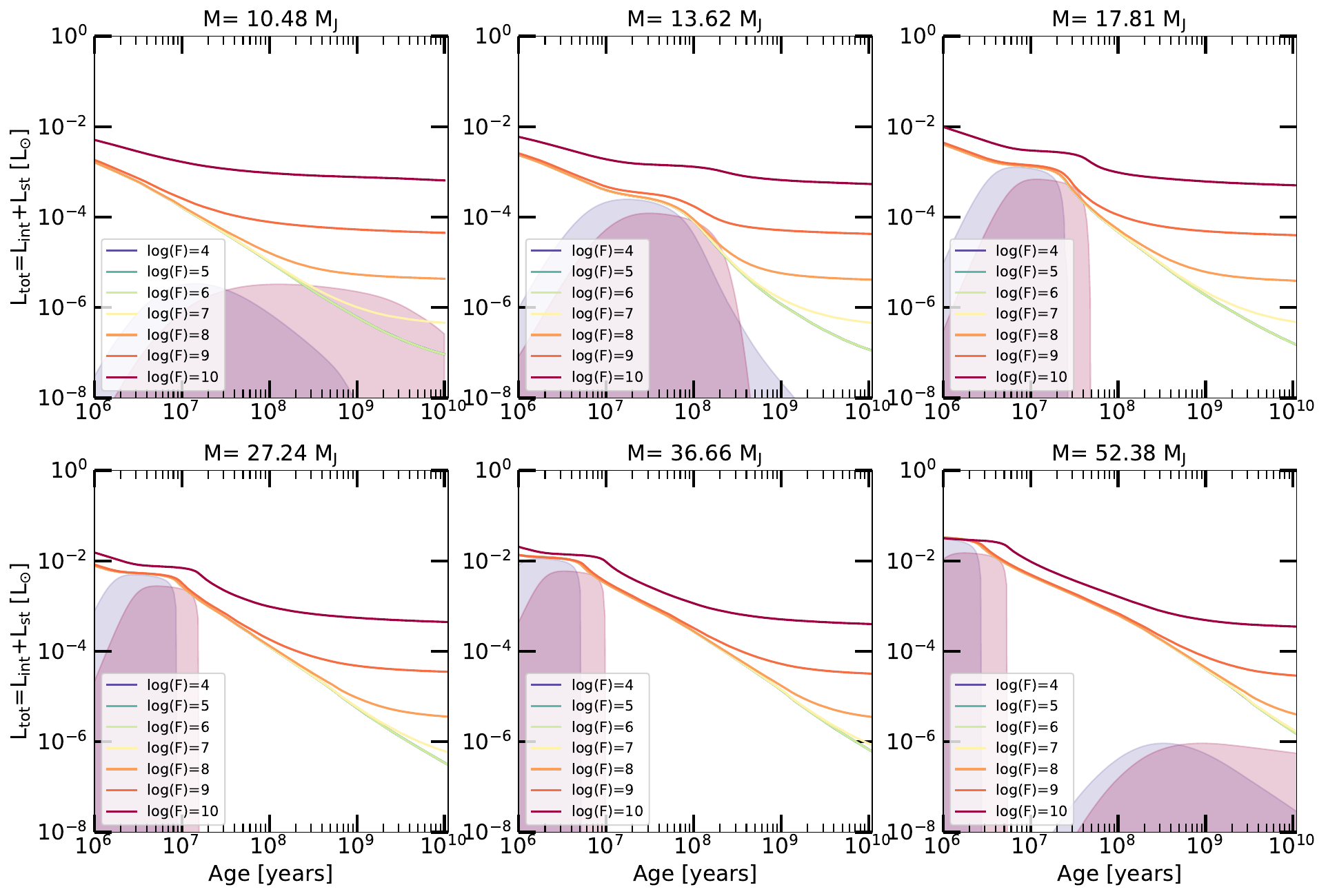}
    \caption{Each panel shows the effect of incident stellar radiation on the evolution of total luminosity emitted by irradiated brown dwarfs with varying mass. Radius evolution for 10.48, 13.62, 17.81, 27.24, 36.66, and 52.38 $M_{\rm J}$ are shown in the six panels. The different colored lines in each panel show the evolution due to different incident stellar flux on the object with $log(F)$ varying from 4 to 10, where $F$ is in $erg.s^{-1}.cm^{-2}$. The nuclear luminosity component for the model with the least amount of incident flux ($log(F)=4$) is shown with the blue shaded region, while the red shaded region shows the same but for the case with $log(F)=10$. Note that the total luminosity plotted here includes the internal luminosity of the object as well as the incident stellar radiation component that is re-radiated to space by the object.}
    \label{fig:lum_w_radius}
\end{figure*}

Figure \ref{fig:TP-profile} shows that incident flux from the host star can cause irradiated brown dwarf atmospheres to be radiative at higher pressures compared to field brown dwarfs. This effect can significantly decrease their rate of cooling leading to larger radii compared to field brown dwarfs for a given mass at a given age. The six panels of Figure \ref{fig:stellar_w_radius} show this effect for transiting brown dwarfs of six different masses between 10--53 $M_{\rm J}$. The different colors in each panel show the radius for a different $log(F)$ between 4-10 in cgs units, which corresponds to a $T_{\rm eq}$ range between $\sim$81-2576 K. Figure \ref{fig:stellar_w_radius} shows that the radius enhancement caused by incident stellar flux becomes a major effect across all masses when the incident flux is higher than $log_{10}(F)\ge9$ or $T_{\rm eq}\ge1450$ K, where $F$ is in $erg.s^{-1}.cm^{-2}$. Figure \ref{fig:mass_radius_stellar} shows the effect of the incident stellar irradiation on the mass--radius relation for transiting brown dwarfs. Mass--radius relations for five different ages between 21.5 Myrs and 10 Gyrs are shown with different colors in Figure \ref{fig:mass_radius_stellar}. The mass-radius relation for varying levels of incident stellar flux is shown with different line styles for each age. The area between the lowest and highest levels of incident stellar flux is shaded for clarity.

A major revelation from Figure \ref{fig:mass_radius_stellar} is that the radius enhancement due to incident stellar flux is largely independent of the object mass at both young and older ages. This result holds true for all masses except for the small fraction of time when their contraction is halted due to deuterium burning phase. This deuterium burning phase is marked by the sharp increase and subsequent decrease in radius with increasing mass at each age in Figure \ref{fig:mass_radius_stellar}. Figure \ref{fig:mass_radius_stellar} also shows that the radius enhancement caused by the incident stellar flux decreases with increasing age for all masses. This can be seen as the shaded areas get narrower with increasing age in Figure \ref{fig:mass_radius_stellar}. For example, for a 31 $M_{\rm J}$ object, its radius inflation between $logF=10$ ($T_{\rm eq}$=2576 K) and $logF=6$ ($T_{\rm eq}$=257 K) is 28\% at 21.5 Myrs, but this inflation is 16\% at 10 Gyrs. This suggests that the a young irradiated brown dwarf should appear more inflated than an isolated brown dwarf (at the same mass and age). The increase in radius (relative to isolated brown dwarfs) due to stellar irradiation would be comparably less in an older system.

Figure \ref{fig:lum_w_radius} shows the luminosity evolution of the same objects for different levels of incident stellar flux. The total emitted luminosity from the object is shown, which includes contributions from the emitted interior heat as well as the incident stellar energy that is remitted back to space by the brown dwarf. The internal luminosity dominates the total emitted luminosity at young ages in these objects. However, the incident stellar irradiation becomes the dominant contributor to the emitted luminosity after a certain age, which depends on the object mass. Figure \ref{fig:lum_w_radius} shows that this transition occurs at younger ages for objects with lower mass. With increasing object mass, the effect of stellar irradiation on the total luminosity budget of these objects decreases. 

The component of the emitted luminosity originating from nuclear reactions in the interior is also shown in each panel of Figure \ref{fig:lum_w_radius} as shaded regions. Only the nuclear luminosities for the cases with the lowest incident stellar flux ($log(F)=4$ in blue) and the highest incident stellar flux ($log(F)=10$ in red) are shown for clarity. Each panel shows that the incident stellar flux significantly alters the emitted nuclear luminosity as well as the timescale associated with it. We discuss these effects and their implications in the following sections.

\subsubsection{Deuterium Burning Minimum Mass}

\begin{figure*}
    \centering
    \includegraphics[width=1\linewidth]{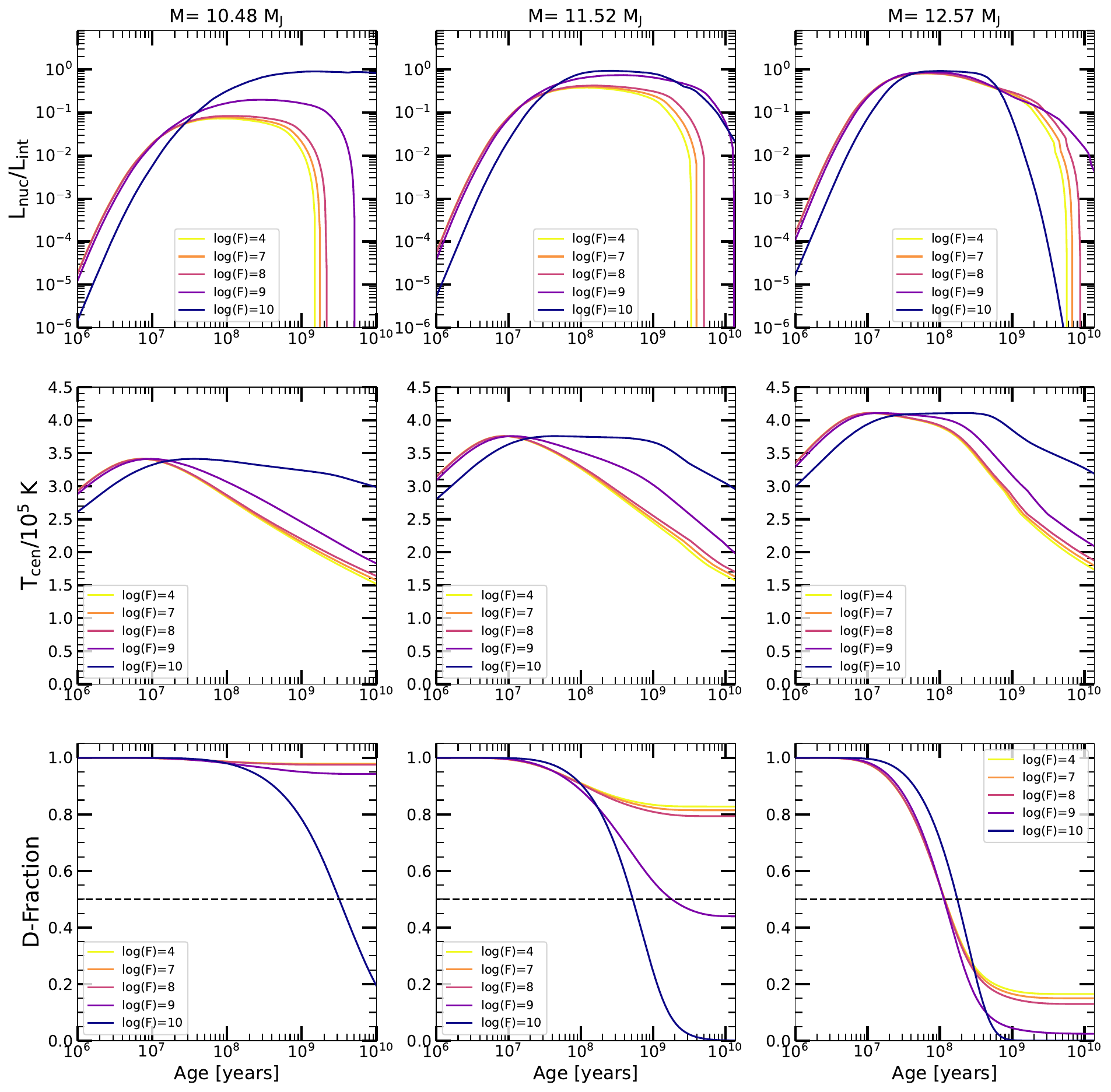}
    \caption{The effect of stellar irradiation on the evolution of irradiated brown dwarfs near the deuterium burning minimum mass is shown. The left, middle, and right columns show the evolution for 10.48, 11.52, and 12.57 $M_{\rm J}$. The top, middle, and bottom row show the nuclear luminosity to internal luminosity fraction, central temperature, and D- fraction with time, respectively. Each panel shows models for varying incident stellar flux. For example, luminosity from deuterium burning comprises almost all of the internal luminosity emitted by a 10.48 $M_{\rm J}$ object that receives $log(F)=10$ but the deuterium burning contributes less than 10\% of the internal luminosity for the same object if it receives $log(F)=7$. This shows that the deuterium burning minimum mass is a function of incident stellar flux for irradiated brown dwarfs.}
    \label{fig:dbmm_detailed}
\end{figure*}

\begin{figure*}
    \centering
    \includegraphics[width=0.45\linewidth]{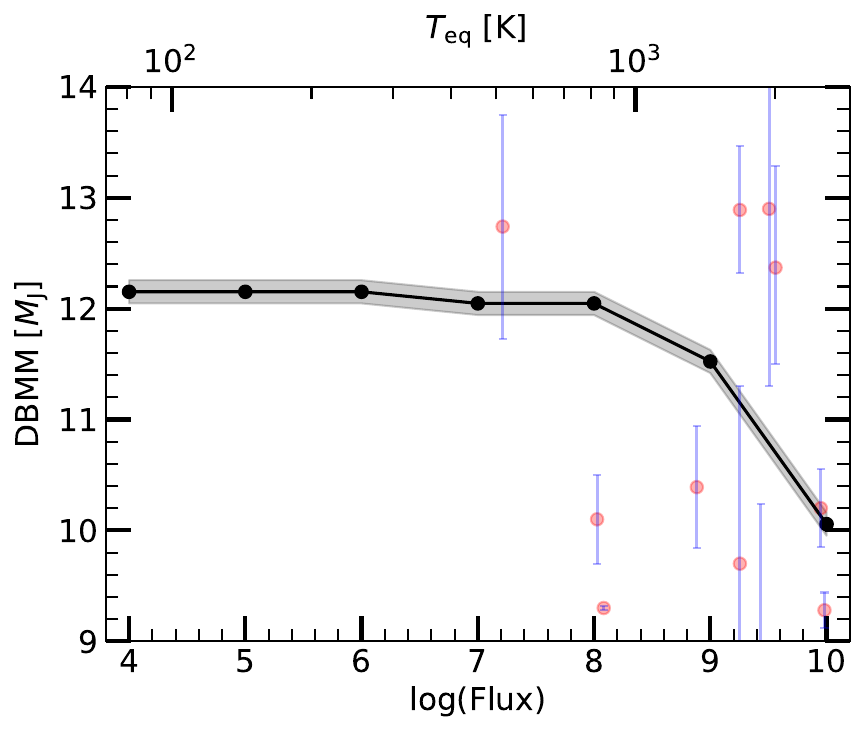}
    \includegraphics[width=0.45\linewidth]{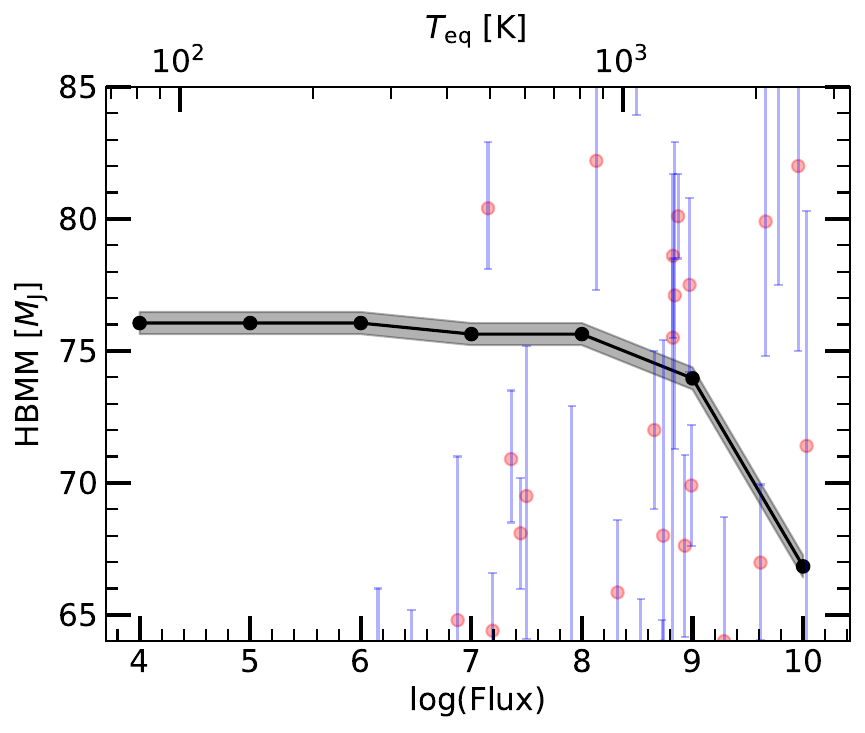}
    \caption{Left panel shows the dependence of the deuterium burning minimum mass on the incident stellar flux while the right panel shows the dependence of the Hydrogen burning minimum mass on the incident stellar flux. The zero-albedo equilibrium temperature for each incident flux is shown in the top x-axis. Sources from the transiting exoplanets and transiting brown dwarf samples are also shown in both panels.}
    \label{fig:dbmm_hbmm}
\end{figure*}

 The deuterium burning minimum mass (DBMM), often assumed to be the mass boundary separating planets and brown dwarfs, is defined as the minimum object mass for which $\ge$50\% of the initial $^2$H inventory of the object is exhausted within the age of the universe \citep[e.g.,][]{diamondback,bodenheimer13}. We explore how stellar irradiation affects this boundary here.  Figure \ref{fig:TP-profile} shows that incident stellar energy causes the radiative--convective boundary to move to higher pressures compared to objects without external irradiation. This thicker radiative zone means that less flux can be transported through the atmosphere at a given log $g$, such that a higher specific entropy adiabat is needed for a converged model with a given specified $T_{\rm int}$.  In a cooling calculation, a given $T_{\rm int}$ is reached at a higher specific entropy of the deep atmosphere (and interior). This leads to higher central temperature in these objects compared to isolated brown dwarfs.

As the rate of nuclear reactions are very sensitive to the central temperature, irradiated brown dwarfs show higher nuclear luminosity relative to isolated brown dwarfs at a given mass, which is shown both in Figure \ref{fig:lum_w_radius} and the top three panels of Figure \ref{fig:dbmm_detailed}, which show the fraction of the interior luminosity contributed by the nuclear luminosity for varying stellar fluxes for three object masses-- 10.48, 11.52, and 12.57 $M_{\rm J}$ in the left, middle, and right panels, respectively. For 10.48 $M_{\rm J}$, deuterium burning contributes less than 10\% of the interior luminosity between 10--1000 Myrs, if the incident flux is lower than $log(F) \le 8$. This fraction is slightly larger and is sustained between 20--4000 Myrs if $log(F)=9$. However, deuterium burning contributes to more than 90\% of the internal luminosity after $\sim$400 Myrs till 10 Gyrs if  $log(F)=10$. The middle row in Figure \ref{fig:dbmm_detailed} shows the central temperatures for all the models shown in the top panels. The significantly higher nuclear luminosity is due to significantly higher central temperature in the $log(F)=10$ case compared to other smaller incident flux models for the 10.48 $M_{\rm J}$ object.

The panels in the bottom row show the fraction of deuterium available in the models shown in the top and middle panels. The bottom left panel shows that more than 80\% of the initial deuterium inventory is exhausted in the $log(F)=10$ model for the 10.48 $M_{\rm J}$ object due to sustained deuterium burning. This fraction is less than 5\% for lower flux levels for the same object mass as their nuclear luminosity contribution to the internal luminosity is an order of magnitude smaller. This suggests that if $log(F)=10$, then a 10.48 $M_{\rm J}$ mass object should be considered above the deuterium burning minimum mass. However, if $log(F) \le 9$ then the same object would be considered below the deuterium burning minimum mass. 

The middle column of Figure \ref{fig:dbmm_detailed} shows that a 11.52 $M_{\rm J}$ can be considered above the deuterium burning minimum mass if $log(F)\ge 9$ but if  $log(F)\le 8$ then such an object is still below this limit. The right column of Figure \ref{fig:dbmm_detailed} shows that a 12.57 $M_{\rm J}$ object is above the deuterium burning minimum mass for all fluxes considered in this work. The left panel of Figure \ref{fig:dbmm_hbmm} shows how the deuterium burning minimum mass varies as a function of the incident stellar flux on transiting substellar objects, showing a decrease with increasing incident flux. The DBMM obtain by \citet{diamondback} for evolution with clear atmospheres were 12.16 $M_{\rm J}$, which matches very well with our calculation for objects with low stellar irradiation. The DBMM obtained from the solar composition ``hybrid" evolutionary models in \citet{diamondback} were slightly lower 12.05 $M_{\rm J}$, which highlights the impact of clouds on the evolution of these objects.  This further suggests that the deuterium burning minimum mass, which is often used as a distinction between planets and brown dwarfs \citep{spiegel11,molliere12,chabrier23,diamondback}, is a rather fuzzy boundary.

\subsubsection{Hydrogen Burning Minimum Mass}

\begin{figure*}
    \centering
    \includegraphics[width=1\linewidth]{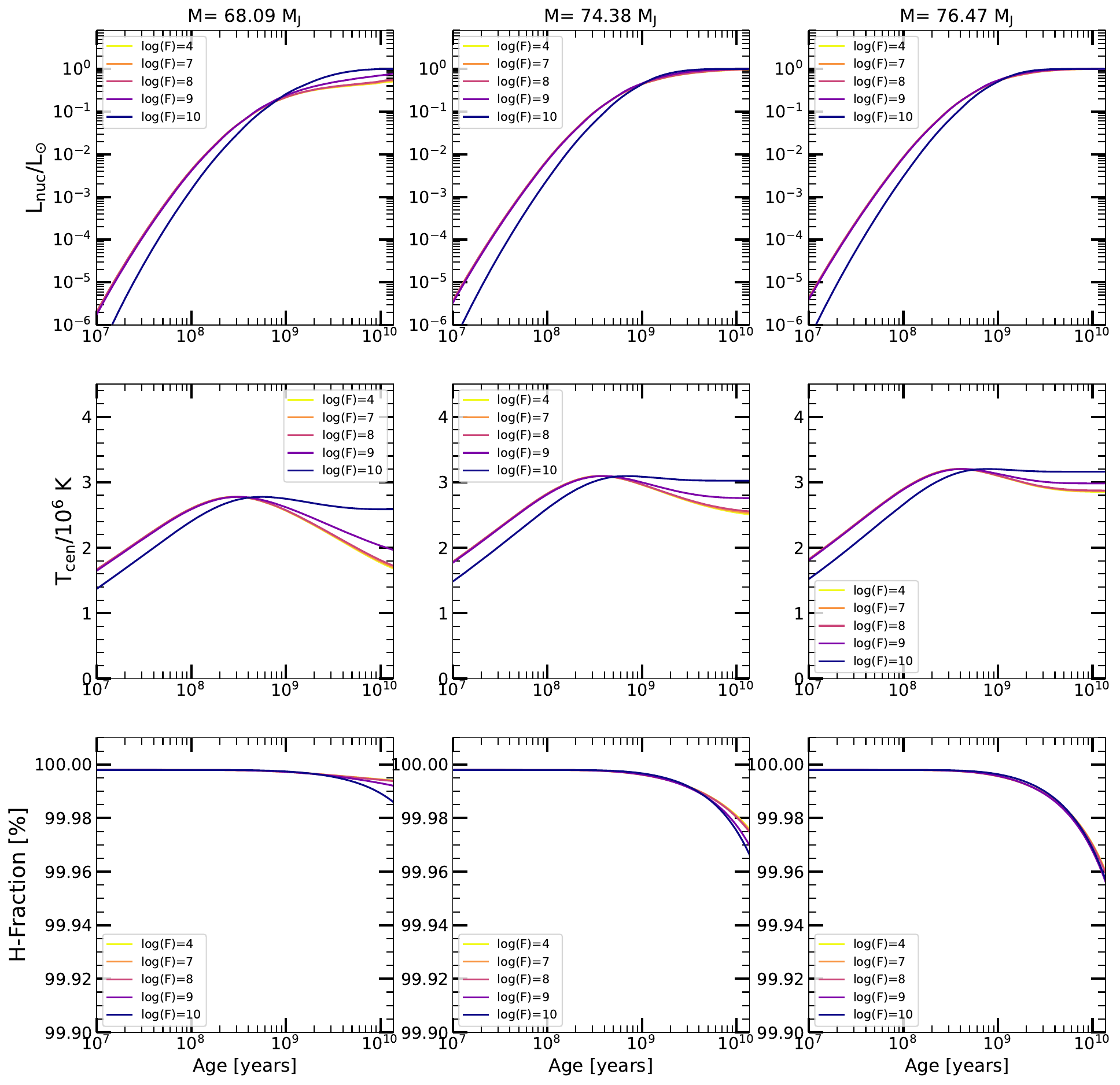}
    \caption{This figure shows the same quantities as Figure \ref{fig:dbmm_detailed} with the exception of the last row, which now shows the H- fraction instead of the D- fraction. The dependance of the evolutionary properties of objects near the Hydrogen burning minimum mass on the incident stellar flux is shown in the three columns here. Left, middle, and right columns show evolution for 68.09, 74.38, and 76.47 $M_{\rm J}$ objects.}
    \label{fig:hbmm_detailed}
\end{figure*}

The effects of instellation also operate in the higher mass end, influencing the H-burning minimum mass. The H-burning minimum mass is defined as the mass above which the nuclear luminosity from H-burning becomes constant with age. The right panel of Figure \ref{fig:dbmm_hbmm} shows the decrease of the H-burning minimum mass as a function of incident stellar flux. A 68 $M_{\rm J}$ is well below the H-burning minimum mass for $log(F) \le 9$ as its nuclear luminosity from H-burning shows decline with age. However, such a low mass object can sustain a constant nuclear luminosity from H-burning if the incident stellar flux is $log(F)\ge 10$. This effect can be seen in Figure \ref{fig:hbmm_detailed}, which is similar to Figure \ref{fig:dbmm_detailed} but for object masses near the H-burning minimum mass regime.

The left column in Figure \ref{fig:hbmm_detailed} shows that a 68 $M_{\rm J}$ object can sustain a constant luminosity and central temperature for 13.7 Gyrs or more if the incident stellar flux is $log(F)\ge 10$. However, such an object fails to meet either of those criterias if $log(F)\le 9$. A constant nuclear luminosity and central temperature can be achieved for a 74.38 $M_{\rm J}$ (middle column in Figure \ref{fig:hbmm_detailed}) object if $log(F)\ge 9$ instead. A 76.47 $M_{\rm J}$ object meets these criteria for all of these incident flux levels. The HBMM from the evolution of clear brown dwarfs with solar composition was found to be 76.5 $M_{\rm J}$ in \citet{diamondback}, which matches well with our calculated values for objects with low incident stellar fluxes. The right column of Figure \ref{fig:dbmm_hbmm} shows the HBMM as a function of incident stellar flux.

\subsection{Effects of Cores}

\begin{figure*}
    \centering
    \includegraphics[width=0.48\linewidth]{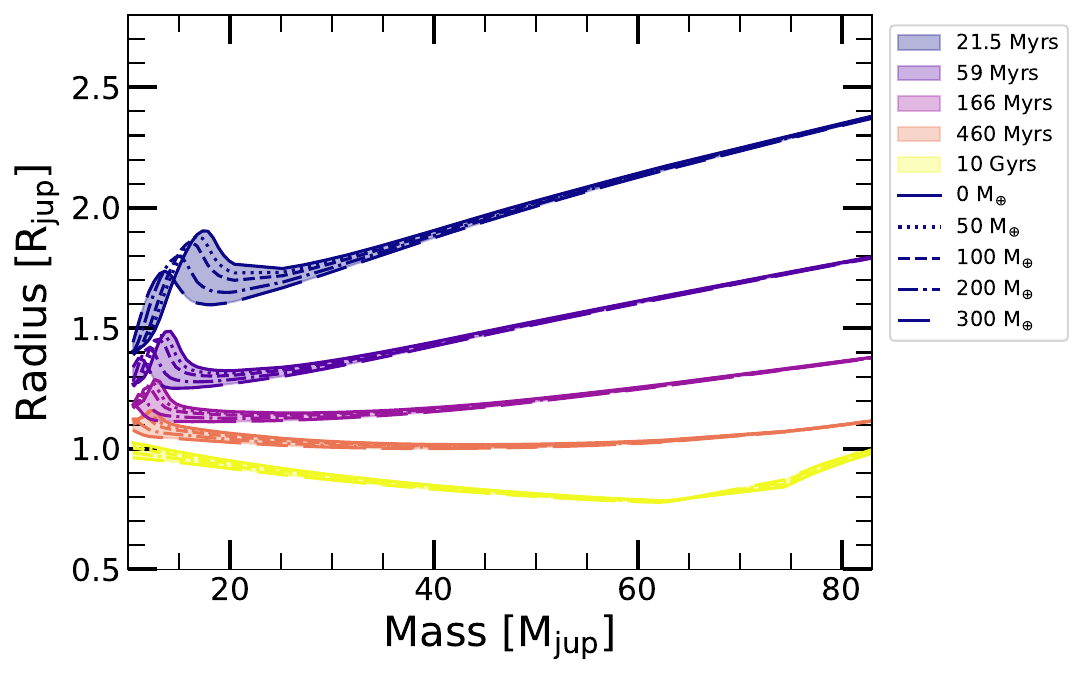}
    \includegraphics[width=0.48\linewidth]{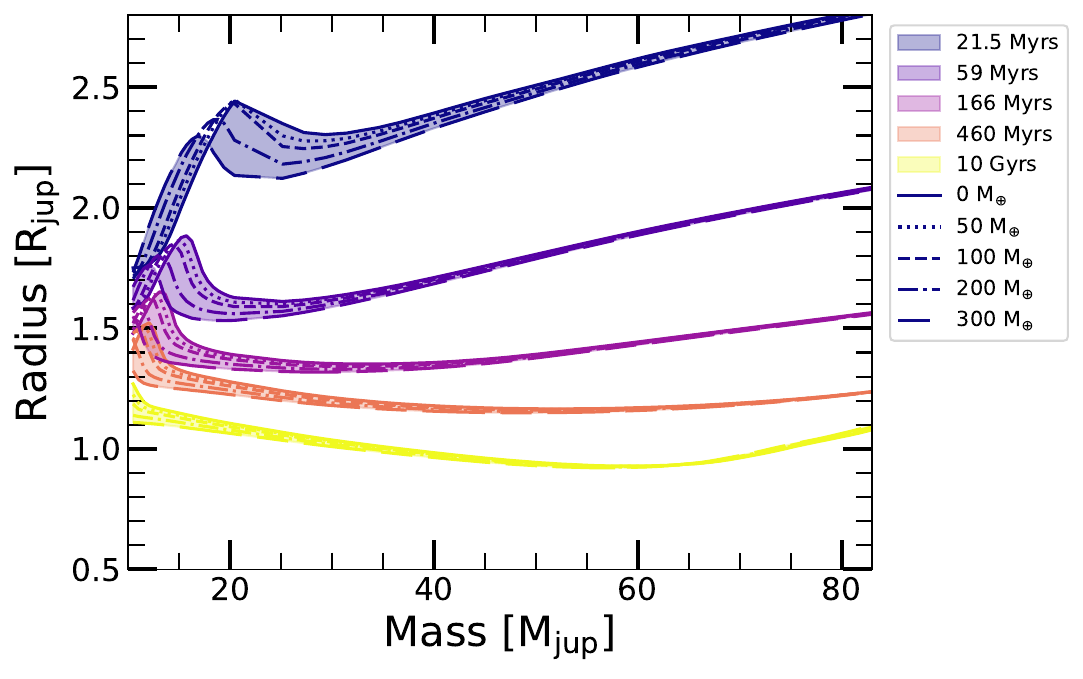}
    \caption{The effect of metal cores of different masses on the mass-radius relation for irradiated brown dwarfs is shown in the two panels here. The left panel shows models when $log(F)=6$ and the right panel shows models for $log(F)=10$. The shaded region around the mass-radius relations at each age show the change in the relation due to changing core masses from 0 to 300 $M_{\oplus}$. The mass-radius relation for 21.5 Myrs, 59 Myrs, 166 Myrs, 460 Myrs, and 10 Gyrs is shown.}
    \label{fig:mass_radius_core}
\end{figure*}

Transiting substellar objects may accrete significantly larger amounts of metals compared to free-floating field brown dwarfs as they are either formed or they evolve in protoplanetary disk environments close to their host stars. These objects may form in a disk via core accretion, and therefore, we investigate the role of metal cores in the structure and evolution of transiting brown dwarfs.

The left and right panels of Figure \ref{fig:mass_radius_core} show the effect of cores on the radius evolution of irradiated brown dwarfs if the incident flux is $log(F)=6$ and $log(F)=10$, respectively. To explore the maximal extent to which metal cores can affect their radius, we have explored core masses-- 0, 50, 100, 200, and 300 $M_{\oplus}$. The equation of state of the cores are implemented as a 50\% mixture of rocks and ice from \citet{ANEOS}. The nuclear burning is also implemented only for mass shells that are external to the core, if the core mass is non-zero. The core masses have the maximum impact on the radius of irradiated brown dwarfs near their D- burning phase. Figure \ref{fig:mass_radius_core} shows that at each age beyond a certain object mass, objects with more massive cores appear smaller in radius compared to objects with less massive cores. For example, if $log(F)=6$ and the age is 21.5 Myrs, objects more massive than $\sim$17 $M_{\rm J}$ follow this trend. However, the trend reverses for $M<$15$M_{\rm J}$, where objects with smaller core masses appear smaller in radius too. This reversal is limited to the D-burning phase of their evolution, where higher core mass causes higher temperatures at the core-envelope interface leading to more nuclear luminosity from D-burning \citep{molliere12}. The reversal no longer appears once the D-burning phase ends. The same trend reversal also occurs for the highly irradiated objects with $log(F)=10$, albeit at a slightly higher object mass than $log(F)=6$. Figure \ref{fig:mass_radius_core} also shows that the presence of such metal cores have a much smaller effect on the radii of transiting brown dwarfs compared to the impact of the incident stellar flux. This is understandable, as the masses of these objects are much larger than the core masses considered here.

\subsection{Migration}

\begin{figure*}
    \centering
    \includegraphics[width=0.9\linewidth]{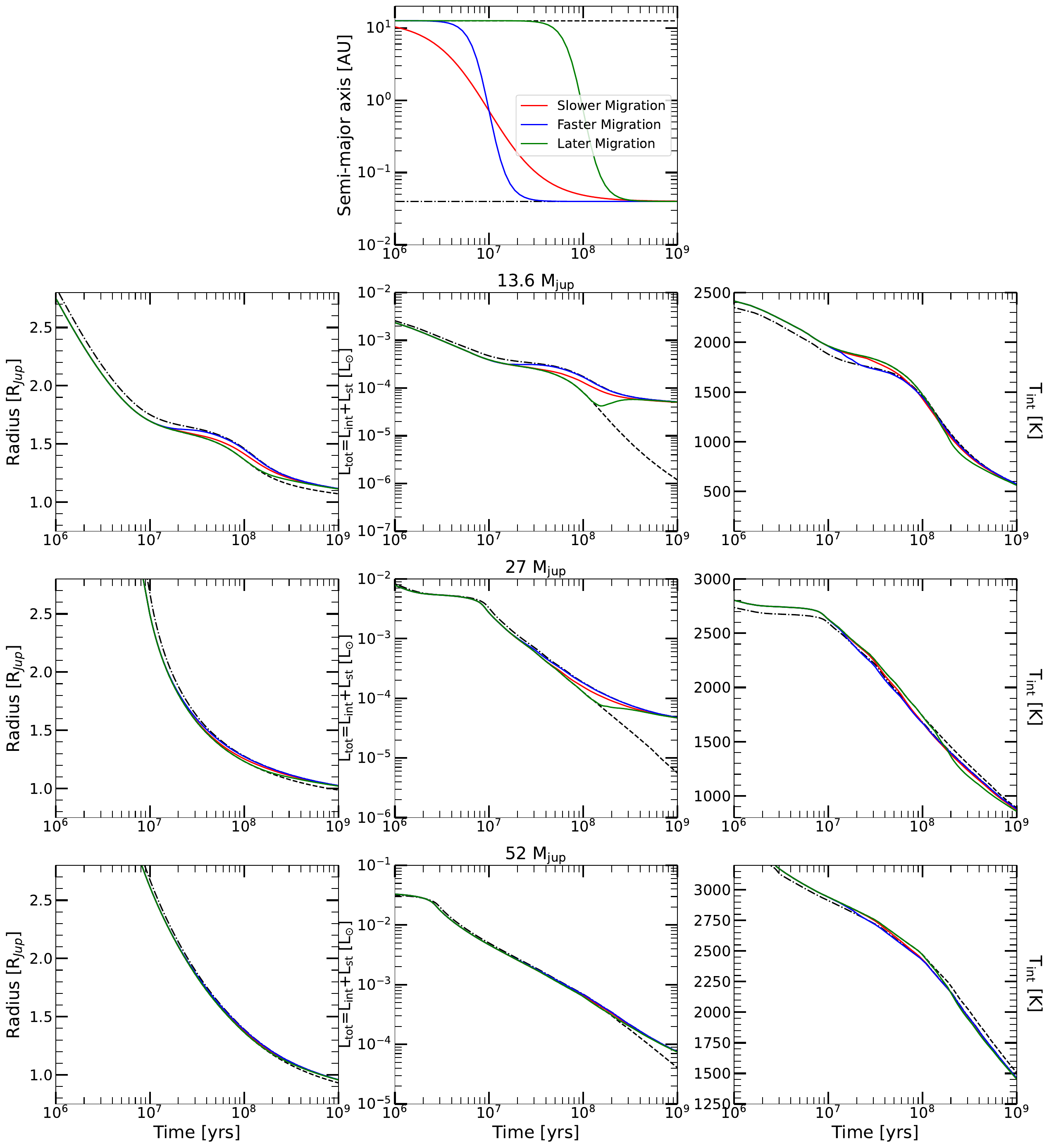}
    \caption{Effect of orbital migration on the evolution of irradiated brown dwarfs is shown. The top panel shows the variation of the semi-major axis with time for the five orbital evolutions considered here-- 1) object stays at 12.6 AU (dashed lines in Figure \ref{fig:migration}), 2) object stays at 0.04 AU (dot-dashed lines in Figure \ref{fig:migration}), 3) object migrates from 12.6 AU to 0.04 AU relatively quickly (blue lines), 4) object migrates between same distances much more slowly (red lines), and 5) object migrates between those distances quickly but starts the migration late (green lines). The second, third, and the last row shows evolution for object mass 13.6, 27, and 52 $M_{\rm J}$. The left, middle, and right columns shows the radius evolution, total luminosity evolution, and $T_{\rm int}$ evolution for these four migration scenarios.}
    \label{fig:migration}
\end{figure*}

Even though transiting brown dwarfs are often found in close proximity to their host stars, significant fractions of their evolutionary lifetimes might be spent migrating inwards from faraway locations in the protoplanetary disks. We have explored the effects of varying incident stellar flux on the object's evolution during such migration.

 We have compared the evolution across five types of migration -- 1) object stays at 12.6 AU (dashed lines in Figure \ref{fig:migration}), 2) object stays at 0.04 AU (dot-dashed lines in Figure \ref{fig:migration}), 3) object migrates from 12.6 AU to 0.04 AU at $\sim$10 Myrs relatively quickly (blue lines), 4) object migrates between same distances at $\sim$10 Myrs slowly (red lines), and 5) late and fast migration (green lines) happens at $\sim$100 Myrs. We have simulated the orbital evolutions for three representative masses - 13.6 $M_{\rm J}$, 27 $M_{\rm J}$, and 52 $M_{\rm J}$. Top most panel of Figure \ref{fig:migration} shows the assumed temporal evolution of the object semi-major axis in each of these five cases. The 2nd, 3rd, and the last row of Figure \ref{fig:migration} show the evolution for 13.6 $M_{\rm J}$, 27 $M_{\rm J}$, and 52 $M_{\rm J}$, respectively. The radius evolution, luminosity evolution, and the $T_{\rm int}$ evolution is shown as a function of time in the left, middle, and right columns. 

Comparing the three rows of Figure \ref{fig:migration} shows that the effect of the migration history on the evolution is much larger for low mass objects than massive objects. Figure \ref{fig:migration} also show that the effect of this migration on the evolution is delayed until the object reaches a certain proximity threshold to their host stars. For example, the semi-major axis continually decreases since before 1 Myrs in the slow migration case (red lines) but the effect of this migration on the object properties only becomes visible for ages beyond 20 Myrs. At this stage, the radius contraction slows down significantly compared to the objects that remained at 12.6 AU. The object reaches its final semi-major axis of 0.04 AU at $\sim$300 Myrs (red line) and at this stage the radius evolution of the migrating object also starts to behave similarly to the object which remained at 0.04 AU. If the migration is significantly faster (blue lines), this change in the radius evolution also occurs quickly. Moreover, the effects of the migration history are not present on the object properties (radius, luminosity, etc) after the migration is complete. If the migration is relatively late and rapid, e.g., if it begins at 100 Myrs, the object still settles on the properties similar to the case of static evolution at 0.04 AU within $\sim$1 Gyrs (green lines). Therefore, our results suggest that migration is unlikely to have a major impact on the properties of transiting brown dwarfs unless it is a still ongoing process. However, we note that orbital migration could affect the evolution of transiting brown dwarfs in additional ways not considered here, e.g. through tides.

\subsection{Anomalous heating from incident flux}

\begin{figure}
    \centering
    \includegraphics[width=1.0\linewidth]{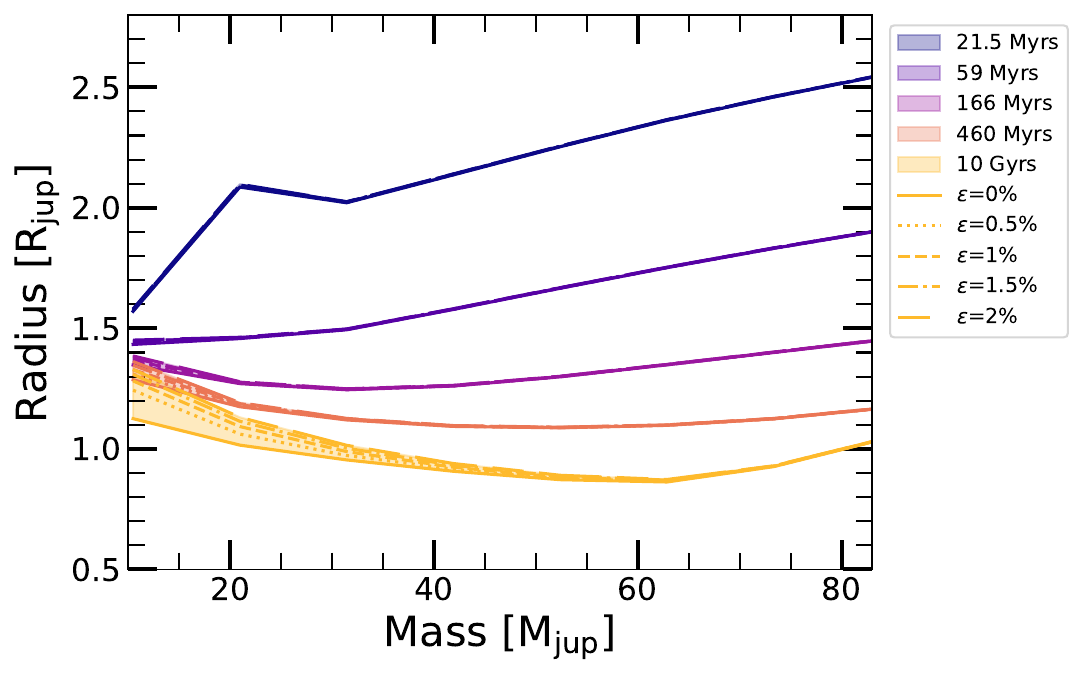}
    \caption{Effect of anomalous heating on the mass-radius relation of irradiated brown dwarfs at various ages is shown. Each colored shaded region shows the variation in the mass-radius relation due to varying heating efficiency ($\epsilon$) of their interiors. The mass-radius relation for ages 21.5 Myrs, 59 Myrs, 166 Myrs, 480 Myrs, and 10 Gyrs is shown. Models with incident stellar flux $log_{10}(F)=9.5$ are shown here.}
    \label{fig:mass_radius_inf}
\end{figure}

Hot Jupiter exoplanets show anomalously large radii. The exact physics driving this anomaly is yet to be determined but plausible explanations include processes like thermal tides or ohmic dissipation heating their interiors \citep{rogers14,Menou12,thorngren24,batygin11,Arras10}. Due to uncertainty regarding the interior heating mechanism, the anomalous heating of hot Jupiter interiors has been modeled using an efficiency parameter $\epsilon$ \citep{komacek17,thorngren2018,Sarkis21}. $\epsilon$ is the fraction of the incident stellar flux that is added to the interior of these hot Jupiters as an anomalous heating term. This parameter has been found to be a strong function of $T_{\rm eq}$ for hot Jupiters and varies within the 2.5\% to 0.2\% range between a $T_{\rm eq}$ range of 1000-2500 K \citep{thorngren2018}. Whether similar heating mechanisms also operate in transiting brown dwarfs is not known yet but here we quantify the extent to which such anomalous heating would change the radii of transiting brown dwarfs. We follow the same formalism as \citet{thorngren2018} to simulate the effects of anomalous heating in transiting brown dwarfs by varying the $\epsilon$ parameter between 0\%-2\%; the $\epsilon=0$ case corresponds to no anomalous heating.

Figure \ref{fig:mass_radius_inf} shows the impact of varying the anomalous heating efficiency parameter on the mass-radius relation for transiting brown dwarfs at various ages. Models with incident stellar flux $log_{10}(F)=9.5$ ($T_{\rm eq}$=1930 K) are shown here. Figure \ref{fig:mass_radius_inf} shows that anomalous heating has little to no effect on the radius of transiting brown dwarfs with ages less than $\sim$ 160 Myrs. However, the effects of anomalous heating start to become important for older low mass objects. At 10 Gyrs, a 10 $M_{\rm J}$ object can have a $\sim$ 24\% larger radius than the case with no anomalous heating, if $\epsilon$=2\%. However, this effect gets weaker with increasing object mass, even at 10 Gyrs. For 10 Gyrs old objects with $M\ge40M_{\rm J}$, the effects of anomalous heating on radius are negligible even if $\epsilon$=2\%. This suggests that older low mass transiting brown dwarfs can serve as a very useful sample to prove/disprove the physical mechanisms that have been proposed for explaining the hot Jupiter radius anomaly problem. We note that the delayed cooling observed in the giant exoplanet population \citep{thorngren2021}, which might be due effects like compositional gradients in the interior or Ohmic dissipation, are not explored in our evolutionary models. We defer a statistical examination of whether the population of transiting brown dwarfs also show delayed cooling till their sample grows larger and the precision on their measured properties, like mass and radius, improve further than the present levels. Instead, we add an additional interior heating term, based on the incident stellar flux, which stalls the evolution of the objects at later ages, keeping their radius inflated. We also note that the internal heating mechanism simulated here is distinct from the radius reinflation mechanism explored in \citet{thorngren2021}, where the effect of an increasing incident flux over time was treated. 

\subsection{Constraints from Observations}

\begin{figure*}
    \centering
    \includegraphics[width=0.3\linewidth]{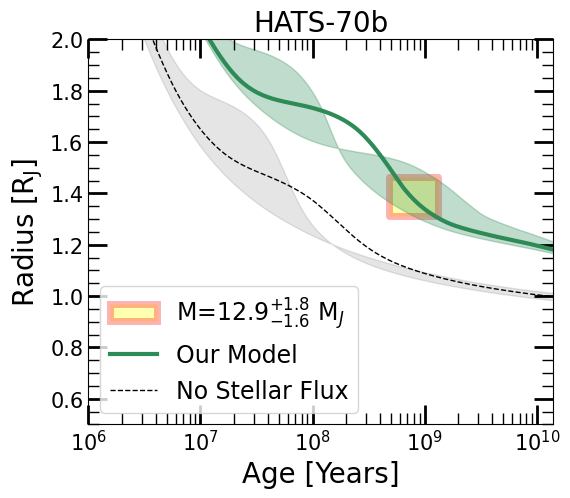}
    \includegraphics[width=0.3\linewidth]{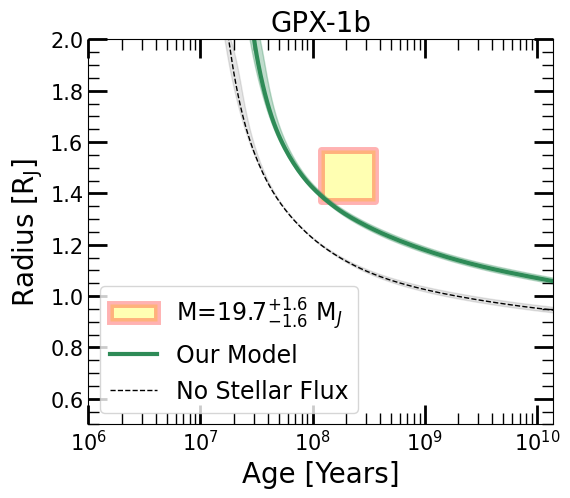}
    \includegraphics[width=0.3\linewidth]{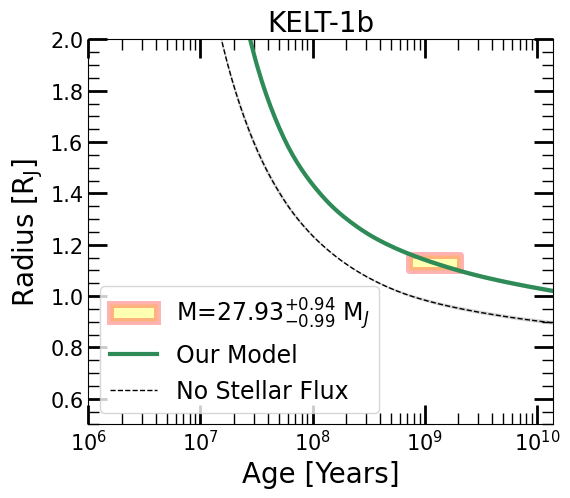}
    \includegraphics[width=0.3\linewidth]{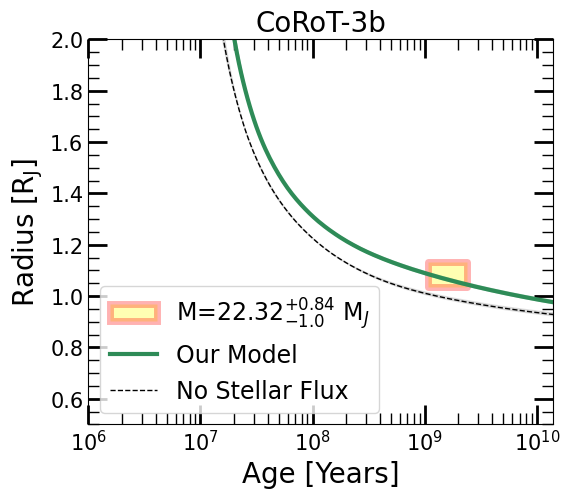}
    \includegraphics[width=0.3\linewidth]{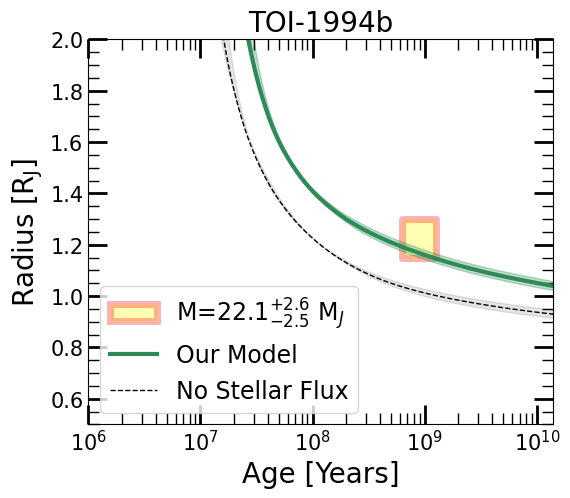}
    \includegraphics[width=0.3\linewidth]{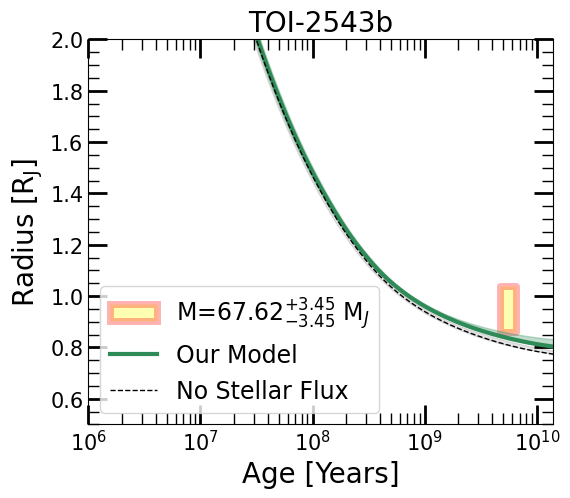}
    \caption{Comparison of our evolutionary models with the measured radii and ages of highly irradiated transiting brown dwarf is shown. The yellow box denotes the measured radius and age and their $1\sigma$ uncertainties. The solid green line shows the irradiated evolutionary model computed for the measured mass $M$, and the green shaded region shows the variation in the evolutionary model due to $1\sigma$ variation on the measured mass. The dashed black line and the gray region show the same but for evolutionary models computed without any stellar irradiation. The evolutionary model that include the effects of stellar irradiation fit the radius of these objects better compared to the model without any stellar irradiation. Note that the green and gray shaded region for HATS-70b do not fully cover the evolutionary track for the median mass for each case because small change in HATS-70b's assumed mass has large effects on its nuclear luminosity from deuterium-burning.}
    \label{fig:fit_obj}
\end{figure*}

\begin{figure*}
    \centering
    \includegraphics[width=1\linewidth]{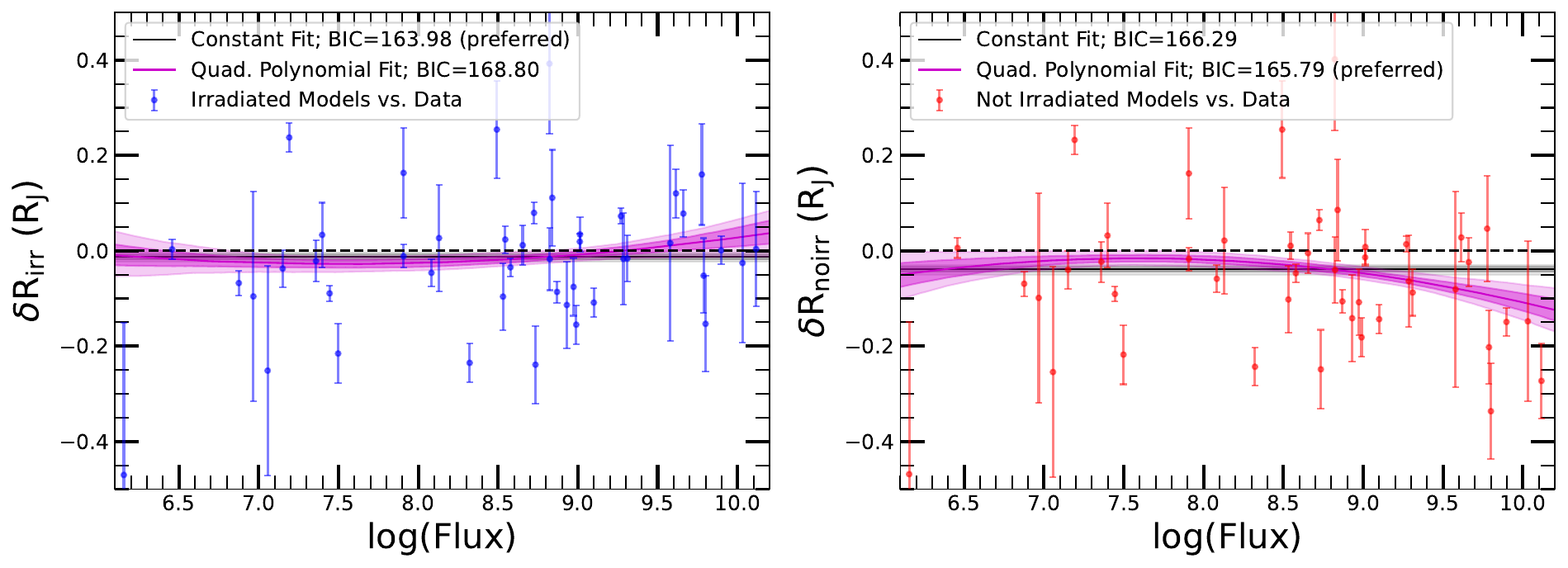}
    \caption{The left panel shows the quantity $\delta{R_{\rm irr}}$ as a function of incident stellar flux for the entire sample of 46 transiting brown dwarfs with masses between 12.9 and 89 $M_{\rm J}$. $\delta{R_{\rm irr}}$ is the difference between the irradiated evolutionary model radius and the measured radius. A positive value indicates that the model over-predicts the radius relative to the measurement, while a negative value indicates the opposite scenario. The right panel shows the same but for the evolutionary models without any stellar irradiation. The magenta line and the shaded regions show the median and $1,2\sigma$ uncertainties on a quadratic polynomial fitted to the trend. The black lines and the shaded region shows the constant function fit to the trends. The BIC values associated with each fit are shown in both the panels. The residuals in the left panel prefer a constant function fit over the quadratic function fit, whereas the residuals in the right panel shows the opposite case. This suggests that the evolutionary models without irradiation fail to match the observed radii of highly irradiated transiting brown dwarfs.}
    \label{fig:fit_better}
\end{figure*}

We use the measured radii, mass, and ages of the current sample of transiting brown dwarfs for comparison with our evolutionary models. We calculate evolutionary models for each object using their mass, host star properties, and semi major axis as inputs. We calculate three evolutionary models for each object-- one corresponding to the measured mass value ($M$), and two other for masses $M+{\delta}M$ and $M-{\delta}M$, where ${\delta}M$ denotes the $1\sigma$ uncertainty on the measured mass. For each of these cases, we also compute evolutionary models with no incident stellar flux. Figure \ref{fig:fit_obj} show the measured radius of 6 transiting brown dwarfs with their radius and age uncertainties denoted with the yellow boxes. The green line and the green shaded region show the radius evolution computed for each object after including the effects of stellar irradiation. The solid green line denotes the model for the reported object mass ($M$) and the shaded green region shows the variation in the radius evolution for $1\sigma$ variation of the object mass. The black dashed line and the gray region show the same for models without any stellar irradiation. For the highly irradiated transiting brown dwarfs shown in Figure \ref{fig:fit_obj}, the evolutionary models that include the effect of stellar irradiation fit the observed radius significantly better than the models without any stellar irradiation.

To present population level trends, we fit the measured radii of 46 objects with masses ranging between 12.9 and 89.0 $M_{\rm J}$ using both the irradiated evolutionary models and the models lacking any incident stellar irradiation. We compute the average model radius within the measured $1\sigma$ age range from each object's evolutionary track. We use the evolutionary track computed with the reported mass $M$ to compute this average radius. We define $\delta{R}$ as the difference between the reported measured radius and this age-averaged model radius. We also define $\sigma_R$=$\sqrt{\sigma_{\rm meas}^2 + \sigma_{\rm mod}^2}$ as the uncertainty on ${\delta}R$, where $\sigma_{\rm meas}$ is the measured radius uncertainty and $\sigma_{\rm mod}$ represents the uncertainty on the model radius. We estimate $\sigma_{\rm mod}$ as the difference between the age-averaged radius between the evolutionary tracks corresponding to the assumed object masses $M+{\delta}M$ and $M-{\delta}M$. 

Figure \ref{fig:fit_better} shows the quantity ${\delta}R$ for the transiting brown dwarf sample as a function of incident flux for the irradiated evolutionary models in the left panel and the models without stellar irradiation in the right panel. A positive value means that the model overestimates the object radius and a negative value suggests the opposite. The right panel in Figure \ref{fig:fit_better} clearly shows that the evolutionary models without irradiation tends to increasingly underestimate the object radius with increasing incident flux for objects with $log(F)>8.5$. The irradiated evolutionary models on the other hand do not show such systematic dependence of ${\delta}R$ with incident flux. Both type of models perform very similarly for lower flux levels than $\sim{log(F)}<8.5$.

 We quantify the extent to which our irradiated evolutionary models better fit the radii of transiting brown dwarfs by fitting the ${\delta}R(F)$ in Figure \ref{fig:fit_better} with two parametric relations -- 1) a constant value with flux, and 2) a quadratic relation with flux. We use PyMultiNest \citep{buchner2016} to fit each parametric function to the measured ${\delta}R(F)$ from the irradiated models and the models without any irradiation. The median fit along with 1 and 2$\sigma$ envelopes for each case are shown in Figure \ref{fig:fit_better}, where magenta shows the quadratic fits and the fits from the constant value are shown in black. We compare the $BIC$ values between the two parametric fits for each sample. For the irradiated models shown in the left panel of Figure \ref{fig:fit_better}, we find that the constant function is preferred over the quadratic polynomial with a $\Delta{BIC}$=4.8. On the other hand, we find that the quadratic polynomial is favored over the constant line with a $\Delta{BIC}$=0.5 for the residuals from the models without irradiation. This suggests that the systematic underestimation of radii due to missing host star flux in the evolution models (right panel) are adequately captured by our irradiated evolution models (left panel).
Therefore, the presence of incident stellar flux within evolutionary calculations is preferred by the sample of transiting brown dwarfs, especially by those that receive high levels of stellar irradiation. We note that about $\sim$20\% of transiting brown dwarfs orbit M-dwarf host stars, that have large uncertainties on stellar radii and ages. These uncertainties make comparisons of their measured radii with evolutionary models challenging. Table \ref{tab:tab_objects} presents an object-by-object comparison of the measured radii with the model radii from irradiated and isolated models for all the 46 sources. We find 10 objects show $>3\sigma$ inconsistency between their measured radii and the age-averaged radii predicted from their irradiated evolutionary models. The age-averaged radii predicted from the evolutionary models are smaller than the measured radii for 7 out of these 10 objects. These objects are marked in Table \ref{tab:tab_objects}. We note that a changing model radius over time is averaged over a large age range for objects with large age uncertainties. This might cause some model radii to appear significantly different than the measured radii (e.g., for TOI-2119b), but this is a result of large uncertainty on age and not the performance of the evolutionary model. We include the model radius range within 1$\sigma$ uncertainty on age in Table \ref{tab:tab_objects} to diagnose these cases.

\section{Discussion}\label{sec:discussion}
Key physical factors that can significantly affect the evolution of transiting brown dwarfs have been explored here but several simplifying assumptions about physical effects that could be potentially important have also been made. We discuss these assumptions and their implications in this section. 
\subsection{Clouds}
Aerosols influence the atmospheres of transiting exoplanets, directly imaged exoplanets, and brown dwarfs \citep[e.g.,][]{inglis24,sing16,hoch25,miles22}. The role of aerosols in their atmospheres is rather complex \citep[e.g.,][]{beatty19}, where aerosol cover can both cool down or heat up various parts of the atmosphere. While aerosols in isolated field brown dwarfs are expected to be primarily composed of mineral clouds formed through condensation or chemical reactions \citep[e.g.,][]{morley14,morley2012neglected,helling08}, the case of highly irradiated transiting objects is more complicated. Intense stellar irradiation can lead to the formation of photochemical hazes in addition to condensate clouds \citep[e.g.,][]{ohno20,gao2017sulfur,kawashima18,steinruek23} in these objects. Moreover, due to their significantly asymmetric irradiation environments, the aerosol cover can be patchy, as already suggested by some recent observations of transiting exoplanets \citep[e.g.,][]{mukherjee25W94Ab,fu25limb}. These physical effects can significantly alter the atmospheric energy balance of transiting brown dwarfs, affecting their radiative-convective boundary and the entropy of their deeper adiabat. These changes could potentially change their evolution substantially \citep[e.g.,][]{diamondback,saumonmarley08}. Due to the lack of atmospheric observations of transiting brown dwarfs, we do not know the nature of aerosols and their physical properties on such highly irradiated massive objects. Therefore, we have not included aerosols in our atmospheric and evolutionary models. However, this remains an important issue that needs to be addressed in future work.

\subsection{Disequilibrium Chemistry}

We assume that the atmospheres of transiting brown dwarfs are in thermochemical equilibrium. Although this might be a valid assumption for the deepest parts of their atmospheres, dynamical and photochemical processes can significantly alter the chemistry of their visible atmospheres \citep[e.g.,][]{moses11,tsai17,tsai21,Zahnle14,zahnle16}. While these processes can influence the observable spectra of such objects much more significantly, their effects on the atmospheric $T(P)$ profile are not negligible \citep[e.g.,][]{Mukherjee25a,welbanks24,agundez25,drummond16}. Including these chemical processes within the atmosphere calculation typically leads to about $\sim$100 K or so corrections in the $T(P)$ relative to the thermal profiles computed assuming thermochemical equilibrium. These corrections have been found to be roughly of this order for colder transiting exoplanets and are likely much smaller for hotter objects. However, the impact of such chemical processes on the radiative-convective boundary and the deeper adiabat still needs to be explored for exoplanets and brown dwarfs alike. Therefore, we have ignored these effects in this work but they need to be addressed in future, preferably with some atmospheric observations of transiting brown dwarfs in hand.

\subsection{Applicability to brown dwarf--white dwarf pairs}

 Brown dwarf around white dwarf hosts are rare but they make up another crucial sample, owing to their favorability for atmospheric characterization, for testing the evolutionary models widely used for substellar objects and exoplanets. While our evolutionary models can be applied to understand the evolution and radius inflation of brown dwarfs around white dwarf hosts that are non-interacting \citep{Casewell20,Casewell20b,Casewell2024,Casewell18}, some caveats need to be noted. The atmosphere boundary conditions used to calculate the evolution of transiting brown dwarfs in our models assume a Sun-like spectra for the host star. Owing to their much higher {$T_{\rm eff}$}, white dwarf hosts can often irradiate their companion brown dwarfs with a very different spectral shape compared to Solar spectra. These hosts will likely emit a much higher fraction of their energy in UV or very blue optical wavelengths. The shape of the host star's spectrum is known to affect the thermal structure of planetary atmospheres \citep{molliere15}, which should ultimately affect their deep atmospheric adiabat. A similar effect might also operate in brown dwarfs around white dwarf hosts \citep[e.g.,][]{lothringer20}, altering their thermal evolution than the models presented here. However, we could not quantify the extent to which the host spectral shape influences the evolution of brown dwarfs because the gaseous opacities in \texttt{PICASO} do not extend to UV wavelengths, which is crucial for modeling this effect. But, comparison of the measured radii of brown dwarfs around white dwarf hosts with the predictions from our irradiated evolutionary models might be an empirical diagnostic of this effect. If our predicted radii show systematically greater or smaller degree of mismatch with the measured radii of these objects as a function of host white dwarf $T_{\rm eff}$, this would indicate that the host star spectral shape does influence their thermal evolution. 

\section{Conclusions}\label{sec:conclusions}
We have developed evolutionary and atmospheric models for an emerging class of substellar objects -- transiting brown dwarfs. These objects are rare compared to the much larger known population of isolated field brown dwarfs. However, it is possible to directly measure the mass and radius of the transiting brown dwarfs using radial velocity and transit observations. Therefore, transiting brown dwarfs are a unique testbed for evolutionary models of giant exoplanets and brown dwarfs. We summarize the key conclusions obtained using our theoretical framework for this class of objects below.

\begin{enumerate}

    \item From radiative-convective thermochemical equilibrium modeling of transiting brown dwarf atmospheres, we find that the radiative-convective boundary of transiting brown dwarfs can be significantly deeper than isolated field brown dwarfs. The convective boundary can be at orders-of magnitude higher pressures if their $T_{\rm eq}$ is greater than their $T_{\rm int}$. 

    \item We find that incident stellar irradiation can significantly enhance the object radius of transiting brown dwarfs if the incident stellar flux on these objects is $log_{10}(F)\ge$9 or $T_{\rm eq}\ge1450$ K, where $F$ is in $erg.s^{-1}.cm^{-2}$.

    \item We also find that the radius enhancement of transiting brown dwarfs due to stellar irradiation is largely independent of object mass, except during the brief deuterium-burning phase of their evolution. However, the radius enhancement of younger irradiated transiting brown dwarfs relative to isolated brown dwarfs should be greater compared to older objects.

    \item We also find that the incident stellar irradiation significantly influences the nuclear burning in transiting brown dwarfs. As a result, the deuterium-burning minimum mass (DBMM), often quoted as the ``planet-brown dwarf" mass boundary, is strongly dependent on the incident stellar radiation. The DBMM for an irradiated brown dwarf with $log_{10}(F)=10$ should be near $\sim$10 $M_{\rm J}$ as opposed to $\sim$12 $M_{\rm J}$ for isolated brown dwarfs.

    \item A similar effect also significantly influences the Hydrogen-burning minimum mass, quoted as the ``star-brown dwarf" mass boundary. The HBMM for strongly irradiated brown dwarfs can be $\sim$13\% smaller in mass compared to field brown dwarfs (66.8 $M_{\rm J}$ for $log(F)=10$ compared to the 76 $M_{\rm J}$ for $log(F)=6$). The influence of stellar irradiation on these ``boundaries" are much more pronounced than the impact of clouds or metallicity.

    \item We show that metal cores have a much smaller effect on the evolution of transiting brown dwarfs than stellar irradiation. The effect of cores is maximal during the deuterium-burning phase. This is expected as the significantly massive cores used in our models include a very small fraction of the object mass of these brown dwarfs.

    \item We find that strongly irradiated low mass transiting brown dwarfs ($M\le40$ $M_{\rm J}$) at old ages can be a great sample to test and distinguish between the various plausible anomalous heating mechanisms proposed to explain the hot Jupiter radius anomaly problem.

    \item We also show that varying irradiation over time caused by possible migration of transiting brown dwarfs has a relatively minor impact on their evolution.

    \item We use our evolutionary models to fit the observed radii of a sample of 46 transiting brown dwarfs. We find that the irradiated evolutionary models fit the measured radii of this population significantly better than the evolutionary model without irradiation for objects with $log_{10}(F)\ge$8.5. Their performances are similar for objects with incident fluxes lower than this value. 

    \item  We find that the measured radii of 10 transiting brown dwarfs (Table \ref{tab:tab_objects}) are still inconsistent at a  $>$3$\sigma$ level with the predicted radii from our evolutionary model. This highlights that physical gaps still remain in our irradiated evolutionary models.

\end{enumerate}

Advancements in our understanding of the interior structure and evolution of substellar objects can only be made with more observations, and transiting brown dwarfs can play a very significant role in this endeavor. While we have focused on understanding the radius evolution of these objects, their atmospheres also need to be studied further, both theoretically and observationally. Observational constraints on the $T_{\rm int}$ and [M/H] of transiting brown dwarfs from atmospheric spectroscopy can help us understand their evolution and interior structure better \citep[e.g.,][]{welbanks24,sing24,fortney20}. The influence of aerosols on the properties of these bodies can only be quantified with atmospheric measurements as well. While better constraints on the radius, mass, and age measurements of transiting brown dwarfs will be very useful, constraints from their atmospheres will also benefit both exoplanet and brown dwarf studies immensely.

\begin{acknowledgments}
SM is supported by the Templeton Theory-Experiment (TEX) Cross Training Fellowship from the Templeton foundation and the Heising-Simons foundation through the 51 Pegasi b postdoctoral fellowship. SM also acknowledges use of the lux supercomputer at UC Santa Cruz, funded by NSF MRI grant AST 1828315. TWC is supported by an NSF MPS-Ascend Postdoctoral Fellowship under award 2316566. SM thanks Dr. Luis Welbanks for important feedback on this work.  We thank Noah Vowell for sharing the catalog of lower mass transiting brown dwarfs and massive exoplanets. We also thank the anonymous referee for very insightful feedback that helped us improve this paper.

\end{acknowledgments}

\begin{contribution}
SM came up with the initial research concept and led the theoretical atmospheric, interior, and evolution modeling with key insights and inputs from JJF. SM led the writing of this manuscipt. TWC provided key insights from observations of transiting brown dwarfs and shaped the application of models to observed properties of transiting brown dwarfs. CED and DPT helped develop the interior and evolution model. All co-authors edited the manuscript.


\end{contribution}

%

\software{\texttt{PICASO} \citep{Mukherjee22,batalha19}, Matplotlib \citep{Hunter:2007}, Jupyter Notebook \citep{Kluyver}, Pandas \citep{reback2020pandas}
          }


\appendix

\section{Results of Fitting the Measured Radii of Transiting Brown Dwarfs}

\begin{table}[]
\centering
\begin{tabular}{|l|l|l|l|l|l|l|l|l|}
\hline
Object & Mass (M$_{\rm J}$) & Age (Gyr) & Radius (R$_{\rm J}$) & $log(F)$ & Irr. R (R$_{\rm J}$) & Irr. ${R}(\Delta{t})$ (R$_{\rm J}$) & Iso. R  (R$_{\rm J}$) & Iso. ${R}(\Delta{t})$ (R$_{\rm J}$) \\
\hline\hline
HATS-70b  &  12.9 $_{- 1.6 }^{+ 1.8 }$ &  0.81 $_{- 0.33 }^{+ 0.5 }$ & 1.384 $_{- 0.074 }^{+ 0.079 }$ &  10.12  &  1.388  &  1.307 - 1.499  &  1.111  &  1.076 - 1.14  \\ 
TOI-1278b  &  18.5 $_{- 0.5 }^{+ 0.5 }$ &  4.5 $_{- 3.1 }^{+ 3.1 }$ & 1.09 $_{- 0.2 }^{+ 0.24 }$ &  6.97  &  0.994  &  0.972 - 1.021  &  0.991  &  0.968 - 1.019  \\ 
Kepler-39b  &  19.0 $_{- 1.3 }^{+ 1.3 }$ &  2.2 $_{- 1.5 }^{+ 2.2 }$ & 1.067 $_{- 0.026 }^{+ 0.028 }$ &  8.08  &  1.021  &  0.989 - 1.055  &  1.008  &  0.979 - 1.047  \\ 
GPX-1b  &  19.7 $_{- 1.6 }^{+ 1.6 }$ &  0.27 $_{- 0.15 }^{+ 0.09 }$ & 1.47 $_{- 0.1 }^{+ 0.1 }$ &  9.8  &  1.317  &  1.26 - 1.386  &  1.133  &  1.085 - 1.193  \\ 
TOI-1994b  &  22.1 $_{- 2.5 }^{+ 2.6 }$ &  0.94 $_{- 0.31 }^{+ 0.33 }$ & 1.22 $_{- 0.071 }^{+ 0.082 }$ &  9.79  &  1.168  &  1.147 - 1.19  &  1.018  &  1.003 - 1.034  \\ 
CoRoT-3b  &  22.32 $_{- 1.0 }^{+ 0.84 }$ &  1.58 $_{- 0.5 }^{+ 0.8 }$ & 1.08 $_{- 0.046 }^{+ 0.051 }$ &  9.31  &  1.064  &  1.045 - 1.083  &  0.992  &  0.977 - 1.007  \\ 
KELT-1b  &  27.93 $_{- 0.99 }^{+ 0.94 }$ &  1.3 $_{- 0.58 }^{+ 0.75 }$ & 1.129 $_{- 0.027 }^{+ 0.031 }$ &  9.9  &  1.13  &  1.1 - 1.162  &  0.98  &  0.953 - 1.0  \\ 
NLTT41135b  &  33.7 $_{- 2.6 }^{+ 2.8 }$ &  10.0 $_{- 3.0 }^{+ 3.0 }$ & 1.13 $_{- 0.17 }^{+ 0.27 }$ &  7.06  &  0.879  &  0.872 - 0.886  &  0.876  &  0.869 - 0.883  \\ 
WASP-128b*  &  39.26 $_{- 1.0 }^{+ 0.88 }$ &  0.95 $_{- 0.63 }^{+ 1.0 }$ & 0.957 $_{- 0.016 }^{+ 0.017 }$ &  9.27  &  1.03  &  0.965 - 1.117  &  0.971  &  0.909 - 1.063  \\ 
CWW89Ab  &  39.21 $_{- 1.1 }^{+ 0.91 }$ &  2.76 $_{- 0.61 }^{+ 0.61 }$ & 0.941 $_{- 0.019 }^{+ 0.019 }$ &  8.58  &  0.906  &  0.899 - 0.914  &  0.894  &  0.886 - 0.902  \\ 
KOI-205b  &  39.7 $_{- 1.8 }^{+ 2.1 }$ &  10.2 $_{- 4.3 }^{+ 2.6 }$ & 0.87 $_{- 0.024 }^{+ 0.022 }$ &  7.91  &  0.858  &  0.849 - 0.868  &  0.853  &  0.843 - 0.864  \\ 
TOI-1406b  &  46.0 $_{- 2.7 }^{+ 2.6 }$ &  3.2 $_{- 1.6 }^{+ 2.2 }$ & 0.858 $_{- 0.027 }^{+ 0.028 }$ &  8.54  &  0.882  &  0.857 - 0.912  &  0.868  &  0.846 - 0.9  \\ 
EPIC212036875b  &  52.3 $_{- 1.9 }^{+ 1.9 }$ &  2.7 $_{- 0.84 }^{+ 0.98 }$ & 0.874 $_{- 0.017 }^{+ 0.017 }$ &  9.01  &  0.893  &  0.87 - 0.909  &  0.86  &  0.843 - 0.881  \\ 
TOI-503b  &  53.7 $_{- 1.2 }^{+ 1.2 }$ &  0.18 $_{- 0.11 }^{+ 0.17 }$ & 1.34 $_{- 0.15 }^{+ 0.26 }$ &  9.58  &  1.356  &  1.151 - 1.62  &  1.259  &  1.072 - 1.492  \\ 
TOI-852b  &  53.7 $_{- 1.3 }^{+ 1.4 }$ &  4.04 $_{- 0.76 }^{+ 0.68 }$ & 0.829 $_{- 0.035 }^{+ 0.037 }$ &  9.01  &  0.863  &  0.856 - 0.871  &  0.837  &  0.83 - 0.845  \\ 
AD3116b  &  54.6 $_{- 6.8 }^{+ 6.8 }$ &  0.617 $_{- 0.017 }^{+ 0.017 }$ & 0.954 $_{- 0.065 }^{+ 0.07 }$ &  7.4  &  0.987  &  0.987 - 0.987  &  0.986  &  0.986 - 0.986  \\ 
RIK72b  &  59.2 $_{- 6.7 }^{+ 6.8 }$ &  0.011 $_{- 0.002 }^{+ 0.002 }$ & 3.1 $_{- 0.31 }^{+ 0.31 }$ &  6.16  &  2.63  &  2.501 - 2.762  &  2.631  &  2.501 - 2.765  \\ 
TOI-811b  &  59.9 $_{- 8.6 }^{+ 13.0 }$ &  0.093 $_{- 0.029 }^{+ 0.061 }$ & 1.262 $_{- 0.062 }^{+ 0.062 }$ &  7.91  &  1.425  &  1.288 - 1.576  &  1.424  &  1.28 - 1.584  \\ 
WASP-30b*  &  61.7 $_{- 1.1 }^{+ 1.5 }$ &  4.3 $_{- 1.6 }^{+ 2.4 }$ & 0.964 $_{- 0.028 }^{+ 0.032 }$ &  9.1  &  0.855  &  0.834 - 0.877  &  0.821  &  0.8 - 0.844  \\ 
TOI-263b  &  61.6 $_{- 4.0 }^{+ 4.0 }$ &  10.0 $_{- 7.0 }^{+ 3.0 }$ & 0.91 $_{- 0.07 }^{+ 0.07 }$ &  8.53  &  0.814  &  0.79 - 0.847  &  0.808  &  0.78 - 0.84  \\ 
LHS6343c  &  62.9 $_{- 2.3 }^{+ 2.3 }$ &  2.0 $_{- 1.0 }^{+ 10.0 }$ & 0.833 $_{- 0.021 }^{+ 0.021 }$ &  6.46  &  0.836  &  0.78 - 0.931  &  0.839  &  0.78 - 0.932  \\ 
TOI-569b*  &  63.8 $_{- 1.0 }^{+ 1.0 }$ &  4.7 $_{- 1.3 }^{+ 1.3 }$ & 0.752 $_{- 0.022 }^{+ 0.022 }$ &  8.72  &  0.832  &  0.819 - 0.846  &  0.816  &  0.803 - 0.831  \\ 
TOI-2119b*  &  64.4 $_{- 2.3 }^{+ 2.2 }$ &  1.17 $_{- 1.15 }^{+ 1.15 }$ & 1.083 $_{- 0.027 }^{+ 0.031 }$ &  7.19  &  1.321  &  0.861 - 2.215  &  1.315  &  0.859 - 2.214  \\ 
CoRoT-15b  &  64.0 $_{- 5.0 }^{+ 4.7 }$ &  2.1 $_{- 1.3 }^{+ 2.0 }$ & 0.943 $_{- 0.073 }^{+ 0.12 }$ &  9.29  &  0.926  &  0.873 - 0.998  &  0.879  &  0.823 - 0.958  \\ 
KOI-415b  &  64.8 $_{- 2.0 }^{+ 6.2 }$ &  8.8 $_{- 2.9 }^{+ 2.8 }$ & 0.86 $_{- 0.024 }^{+ 0.026 }$ &  6.88  &  0.792  &  0.781 - 0.805  &  0.791  &  0.78 - 0.803  \\ 
TOI-1982b*  &  65.85 $_{- 2.72 }^{+ 2.75 }$ &  3.6 $_{- 1.5 }^{+ 1.5 }$ & 1.08 $_{- 0.04 }^{+ 0.04 }$ &  8.32  &  0.845  &  0.819 - 0.873  &  0.837  &  0.811 - 0.866  \\ 
TOI-629b  &  66.98 $_{- 2.95 }^{+ 2.96 }$ &  0.32 $_{- 0.13 }^{+ 0.13 }$ & 1.11 $_{- 0.05 }^{+ 0.05 }$ &  9.61  &  1.23  &  1.133 - 1.343  &  1.138  &  1.06 - 1.246  \\ 
TOI-2543b  &  67.62 $_{- 3.45 }^{+ 3.45 }$ &  5.6 $_{- 0.9 }^{+ 0.9 }$ & 0.95 $_{- 0.09 }^{+ 0.09 }$ &  8.93  &  0.836  &  0.83 - 0.843  &  0.809  &  0.802 - 0.815  \\ 
HIP33609b  &  68.0 $_{- 7.1 }^{+ 7.4 }$ &  0.153 $_{- 0.024 }^{+ 0.024 }$ & 1.58 $_{- 0.07 }^{+ 0.074 }$ &  8.73  &  1.341  &  1.298 - 1.385  &  1.331  &  1.288 - 1.376  \\ 
LP261-75b*  &  68.09 $_{- 2.1 }^{+ 2.1 }$ &  10.0 $_{- 7.0 }^{+ 3.0 }$ & 0.898 $_{- 0.015 }^{+ 0.015 }$ &  7.44  &  0.809  &  0.78 - 0.845  &  0.807  &  0.778 - 0.844  \\ 
NGTS-19b*  &  69.5 $_{- 5.4 }^{+ 5.7 }$ &  8.5 $_{- 6.0 }^{+ 3.2 }$ & 1.034 $_{- 0.053 }^{+ 0.055 }$ &  7.5  &  0.818  &  0.787 - 0.864  &  0.816  &  0.784 - 0.859  \\ 
TOI-2336b*  &  69.9 $_{- 2.3 }^{+ 2.3 }$ &  2.4 $_{- 0.5 }^{+ 0.5 }$ & 1.05 $_{- 0.04 }^{+ 0.04 }$ &  8.99  &  0.895  &  0.883 - 0.908  &  0.868  &  0.857 - 0.881  \\ 
EPIC201702477b  &  70.9 $_{- 2.4 }^{+ 2.6 }$ &  9.7 $_{- 4.9 }^{+ 2.9 }$ & 0.83 $_{- 0.036 }^{+ 0.04 }$ &  7.36  &  0.809  &  0.792 - 0.83  &  0.807  &  0.788 - 0.828  \\ 
CoRoT-34b  &  71.4 $_{- 8.6 }^{+ 8.9 }$ &  1.09 $_{- 0.21 }^{+ 0.19 }$ & 1.09 $_{- 0.16 }^{+ 0.17 }$ &  10.03  &  1.064  &  1.048 - 1.083  &  0.942  &  0.925 - 0.96  \\ 
TOI-2533b  &  72.0 $_{- 3.0 }^{+ 3.0 }$ &  4.26 $_{- 1.74 }^{+ 2.18 }$ & 0.85 $_{- 0.03 }^{+ 0.04 }$ &  8.65  &  0.862  &  0.843 - 0.884  &  0.845  &  0.824 - 0.871  \\ 
NGTS-7Ab  &  75.5 $_{- 13.0 }^{+ 3.0 }$ &  0.055 $_{- 0.03 }^{+ 0.08 }$ & 1.38 $_{- 0.13 }^{+ 0.14 }$ &  8.82  &  1.772  &  1.422 - 2.191  &  1.782  &  1.425 - 2.21  \\ 
TOI-148b  &  77.1 $_{- 5.8 }^{+ 5.8 }$ &  7.7 $_{- 3.7 }^{+ 3.7 }$ & 0.81 $_{- 0.06 }^{+ 0.06 }$ &  8.84  &  0.921  &  0.921 - 0.925  &  0.896  &  0.895 - 0.901  \\ 
TOI-2521b  &  77.5 $_{- 3.3 }^{+ 3.3 }$ &  10.1 $_{- 1.1 }^{+ 1.1 }$ & 1.01 $_{- 0.04 }^{+ 0.04 }$ &  8.97  &  0.934  &  0.934 - 0.934  &  0.902  &  0.902 - 0.902  \\ 
Kepler-503b  &  78.6 $_{- 3.1 }^{+ 3.1 }$ &  6.7 $_{- 0.9 }^{+ 1.0 }$ & 0.96 $_{- 0.04 }^{+ 0.06 }$ &  8.82  &  0.943  &  0.943 - 0.943  &  0.92  &  0.92 - 0.92  \\ 
TOI-587b  &  79.9 $_{- 5.1 }^{+ 5.3 }$ &  0.2 $_{- 0.1 }^{+ 0.1 }$ & 1.377 $_{- 0.04 }^{+ 0.043 }$ &  9.66  &  1.455  &  1.281 - 1.658  &  1.353  &  1.196 - 1.534  \\ 
ZTFJ2020+5033*  &  80.1 $_{- 1.6 }^{+ 1.6 }$ &  10.0 $_{- 7.0 }^{+ 3.0 }$ & 1.05 $_{- 0.01 }^{+ 0.01 }$ &  8.87  &  0.964  &  0.964 - 0.966  &  0.944  &  0.944 - 0.946  \\ 
KOI-189b  &  80.4 $_{- 2.3 }^{+ 2.5 }$ &  6.4 $_{- 4.1 }^{+ 4.5 }$ & 0.988 $_{- 0.021 }^{+ 0.02 }$ &  7.15  &  0.951  &  0.95 - 0.955  &  0.948  &  0.948 - 0.953  \\ 
TOI-1712b*  &  82.0 $_{- 7.0 }^{+ 7.0 }$ &  1.62 $_{- 0.03 }^{+ 0.05 }$ & 1.74 $_{- 0.07 }^{+ 0.08 }$ &  9.96  &  1.075  &  1.075 - 1.075  &  0.982  &  0.982 - 0.982  \\ 
TOI-746b  &  82.2 $_{- 4.9 }^{+ 4.9 }$ &  6.5 $_{- 3.9 }^{+ 4.3 }$ & 0.95 $_{- 0.09 }^{+ 0.09 }$ &  8.13  &  0.977  &  0.976 - 0.979  &  0.971  &  0.971 - 0.973  \\ 
EBLMJ0555-57  &  87.9 $_{- 3.98 }^{+ 3.98 }$ &  1.6 $_{- 1.2 }^{+ 1.2 }$ & 0.821 $_{- 0.058 }^{+ 0.128 }$ &  8.49  &  1.075  &  1.054 - 1.17  &  1.075  &  1.047 - 1.163  \\ 
OGLE-TR-123  &  89.0 $_{- 11.5 }^{+ 11.5 }$ &  0.3 $_{- 0.2 }^{+ 0.2 }$ & 1.294 $_{- 0.088 }^{+ 0.088 }$ &  9.78  &  1.454  &  1.226 - 1.757  &  1.34  &  1.146 - 1.599  \\

\hline\hline                                        
\end{tabular}
\caption{The measured mass and radius of the sample of 46 transiting brown dwarfs is shown here along with the calculated radius from the model with stellar irradiation and models without any stellar irradiation. Both the age-averaged radii and the range of model radii within 1$\sigma$ uncertainty on age from irradiated and isolated models are listed.  Objects for which the measured radii show $>3\sigma$ inconsistency with the age-averaged radii predicted from our irradiated evolutionary model are marked with asterisks. The logarithm of incident flux are reported in cgs units.}\label{tab:tab_objects}
\end{table}



\bibliography{sample701, TransitingBDs}{}

\begin{thebibliography}{}
\expandafter\ifx\csname natexlab\endcsname\relax\def\natexlab#1{#1}\fi
\providecommand{\url}[1]{\href{#1}{#1}}
\providecommand{\dodoi}[1]{doi:~\href{http://doi.org/#1}{\nolinkurl{#1}}}
\providecommand{\doeprint}[1]{\href{http://ascl.net/#1}{\nolinkurl{http://ascl.net/#1}}}
\providecommand{\doarXiv}[1]{\href{https://arxiv.org/abs/#1}{\nolinkurl{https://arxiv.org/abs/#1}}}

\bibitem[{L. {Acu{\~n}a-Aguirre} {et~al.}(2025){Acu{\~n}a-Aguirre}, {Kreidberg}, {Molli{\`e}re}, \& {Bachmann}}]{acuna25}
{Acu{\~n}a-Aguirre}, L., {Kreidberg}, L., {Molli{\`e}re}, P., \& {Bachmann}, N. 2025, \bibinfo{title}{{The bulk metal content of WASP-80 b from joint interior-atmosphere retrievals: Breaking degeneracies and exploring biases with panchromatic spectra},} arXiv e-prints, arXiv:2511.13483.
\newblock \doarXiv{2511.13483}

\bibitem[{M. {Ag{\'u}ndez}(2025){Ag{\'u}ndez}}]{agundez25}
{Ag{\'u}ndez}, M. 2025, \bibinfo{title}{{The mutual influence of disequilibrium composition and temperature in exoplanet atmospheres},} \aap, 699, A306, \dodoi{10.1051/0004-6361/202554732}

\bibitem[{D.~S. {Amundsen} {et~al.}(2017){Amundsen}, {Tremblin}, {Manners}, {Baraffe}, \& {Mayne}}]{amundsen17}
{Amundsen}, D.~S., {Tremblin}, P., {Manners}, J., {Baraffe}, I., \& {Mayne}, N.~J. 2017, \bibinfo{title}{{Treatment of overlapping gaseous absorption with the correlated-k method in hot Jupiter and brown dwarf atmosphere models},} \aap, 598, A97, \dodoi{10.1051/0004-6361/201629322}

\bibitem[{P. {Arras} \& A. {Socrates}(2010){Arras} \& {Socrates}}]{Arras10}
{Arras}, P., \& {Socrates}, A. 2010, \bibinfo{title}{{Thermal Tides in Fluid Extrasolar Planets},} \apj, 714, 1, \dodoi{10.1088/0004-637X/714/1/1}

\bibitem[{I. {Baraffe} {et~al.}(2015){Baraffe}, {Homeier}, {Allard}, \& {Chabrier}}]{baraffe15}
{Baraffe}, I., {Homeier}, D., {Allard}, F., \& {Chabrier}, G. 2015, \bibinfo{title}{{New evolutionary models for pre-main sequence and main sequence low-mass stars down to the hydrogen-burning limit},} \aap, 577, A42, \dodoi{10.1051/0004-6361/201425481}

\bibitem[{K. {Barkaoui} {et~al.}(2025){Barkaoui}, {Sebastian}, {Z{\'u}{\~n}iga-Fern{\'a}ndez}, {Triaud}, {Rackham}, {Burgasser}, {Carmichael}, {Gillon}, {Theissen}, {Softich}, {Rojas-Ayala}, {Srdoc}, {Soubkiou}, {Fukui}, {Timmermans}, {Stalport}, {Burdanov}, {Ciardi}, {Collins}, {Davis}, {Davoudi}, {de Wit}, {Demory}, {Deveny}, {Dransfield}, {Ducrot}, {Florian}, {Gan}, {G{\'o}mez Maqueo Chew}, {Hooton}, {Howell}, {Jenkins}, {Littlefield}, {Mart{\'\i}n}, {Murgas}, {Niraula}, {Palle}, {Pedersen}, {Pozuelos}, {Queloz}, {Ricker}, {Schwarz}, {Seager}, {Shporer}, {Scott}, {Stockdale}, \& {Winn}}]{Barkaoui25}
{Barkaoui}, K., {Sebastian}, D., {Z{\'u}{\~n}iga-Fern{\'a}ndez}, S., {et~al.} 2025, \bibinfo{title}{{TOI-6508 b: A massive transiting brown dwarf orbiting a low-mass star},} \aap, 696, A44, \dodoi{10.1051/0004-6361/202453508}

\bibitem[{N.~E. {Batalha} {et~al.}(2019){Batalha}, {Marley}, {Lewis}, \& {Fortney}}]{batalha19}
{Batalha}, N.~E., {Marley}, M.~S., {Lewis}, N.~K., \& {Fortney}, J.~J. 2019, \bibinfo{title}{{Exoplanet Reflected-light Spectroscopy with PICASO},} \apj, 878, 70, \dodoi{10.3847/1538-4357/ab1b51}

\bibitem[{K. {Batygin} {et~al.}(2011){Batygin}, {Stevenson}, \& {Bodenheimer}}]{batygin11}
{Batygin}, K., {Stevenson}, D.~J., \& {Bodenheimer}, P.~H. 2011, \bibinfo{title}{{Evolution of Ohmically Heated Hot Jupiters},} \apj, 738, 1, \dodoi{10.1088/0004-637X/738/1/1}

\bibitem[{T.~G. {Beatty} {et~al.}(2019){Beatty}, {Marley}, {Gaudi}, {Col{\'o}n}, {Fortney}, \& {Showman}}]{beatty19}
{Beatty}, T.~G., {Marley}, M.~S., {Gaudi}, B.~S., {et~al.} 2019, \bibinfo{title}{{Spitzer Phase Curves of KELT-1b and the Signatures of Nightside Clouds in Thermal Phase Observations},} \aj, 158, 166, \dodoi{10.3847/1538-3881/ab33fc}

\bibitem[{T.~G. {Beatty} {et~al.}(2018){Beatty}, {Morley}, {Curtis}, {Burrows}, {Davenport}, \& {Montet}}]{beatty18}
{Beatty}, T.~G., {Morley}, C.~V., {Curtis}, J.~L., {et~al.} 2018, \bibinfo{title}{{A Significant Overluminosity in the Transiting Brown Dwarf CWW 89Ab},} \aj, 156, 168, \dodoi{10.3847/1538-3881/aad697}

\bibitem[{T.~G. {Beatty} {et~al.}(2014){Beatty}, {Collins}, {Fortney}, {Knutson}, {Gaudi}, {Bruns}, {Showman}, {Eastman}, {Pepper}, {Siverd}, {Stassun}, \& {Kielkopf}}]{beatty14}
{Beatty}, T.~G., {Collins}, K.~A., {Fortney}, J., {et~al.} 2014, \bibinfo{title}{{Spitzer and z' Secondary Eclipse Observations of the Highly Irradiated Transiting Brown Dwarf KELT-1b},} \apj, 783, 112, \dodoi{10.1088/0004-637X/783/2/112}

\bibitem[{S.~A. {Beiler} {et~al.}(2024){Beiler}, {Cushing}, {Kirkpatrick}, {Schneider}, {Mukherjee}, {Marley}, {Marocco}, \& {Smart}}]{beiler24}
{Beiler}, S.~A., {Cushing}, M.~C., {Kirkpatrick}, J.~D., {et~al.} 2024, \bibinfo{title}{{Precise Bolometric Luminosities and Effective Temperatures of 23 Late-T and Y Dwarfs Obtained with JWST},} \apj, 973, 107, \dodoi{10.3847/1538-4357/ad6301}

\bibitem[{P. {Benni} {et~al.}(2021){Benni}, {Burdanov}, {Krushinsky}, {Bonfanti}, {H{\'e}brard}, {Almenara}, {Dalal}, {Demangeon}, {Tsantaki}, {Pepper}, {Stassun}, {Vanderburg}, {Belinski}, {Kashaev}, {Barkaoui}, {Kim}, {Kang}, {Antonyuk}, {Dyachenko}, {Rastegaev}, {Beskakotov}, {Mitrofanova}, {Pozuelos}, {Kuznetsov}, {Popov}, {Kiefer}, {Wilson}, {Ricker}, {Vanderspek}, {Latham}, {Seager}, {Jenkins}, {Sokov}, {Sokova}, {Marchini}, {Papini}, {Salvaggio}, {Banfi}, {Ba{\c{s}}t{\"u}rk}, {Torun}, {Yal{\c{c}}{\i}nkaya}, {Ivanov}, {Valyavin}, {Jehin}, {Gillon}, {Pak{\v{s}}tien{\.{e}}}, {Hentunen}, {Shadick}, {Bretton}, {W{\"u}nsche}, {Garlitz}, {Jongen}, {Molina}, {Girardin}, {Grau Horta}, {Naves}, {Benkhaldoun}, {Joner}, {Spencer}, {Bieryla}, {Stevens}, {Jensen}, {Collins}, {Charbonneau}, {Quintana}, {Mullally}, \& {Henze}}]{Benni21}
{Benni}, P., {Burdanov}, A.~Y., {Krushinsky}, V.~V., {et~al.} 2021, \bibinfo{title}{{Discovery of a young low-mass brown dwarf transiting a fast-rotating F-type star by the Galactic Plane eXoplanet (GPX) survey},} \mnras, 505, 4956, \dodoi{10.1093/mnras/stab1567}

\bibitem[{P. {Bodenheimer} {et~al.}(2013){Bodenheimer}, {D'Angelo}, {Lissauer}, {Fortney}, \& {Saumon}}]{bodenheimer13}
{Bodenheimer}, P., {D'Angelo}, G., {Lissauer}, J.~J., {Fortney}, J.~J., \& {Saumon}, D. 2013, \bibinfo{title}{{Deuterium Burning in Massive Giant Planets and Low-mass Brown Dwarfs Formed by Core-nucleated Accretion},} \apj, 770, 120, \dodoi{10.1088/0004-637X/770/2/120}

\bibitem[{J. {Buchner}(2016){Buchner}}]{buchner2016}
{Buchner}, J. 2016, {PyMultiNest: Python interface for MultiNest},, Astrophysics Source Code Library, record ascl:1606.005 \doeprint{1606.005}

\bibitem[{A. {Burrows} {et~al.}(2011){Burrows}, {Heng}, \& {Nampaisarn}}]{burrows11}
{Burrows}, A., {Heng}, K., \& {Nampaisarn}, T. 2011, \bibinfo{title}{{The Dependence of Brown Dwarf Radii on Atmospheric Metallicity and Clouds: Theory and Comparison with Observations},} \apj, 736, 47, \dodoi{10.1088/0004-637X/736/1/47}

\bibitem[{A. {Burrows} {et~al.}(2007){Burrows}, {Hubeny}, {Budaj}, \& {Hubbard}}]{burrows07}
{Burrows}, A., {Hubeny}, I., {Budaj}, J., \& {Hubbard}, W.~B. 2007, \bibinfo{title}{{Possible Solutions to the Radius Anomalies of Transiting Giant Planets},} \apj, 661, 502, \dodoi{10.1086/514326}

\bibitem[{A. {Burrows} {et~al.}(1997){Burrows}, {Marley}, {Hubbard}, {Lunine}, {Guillot}, {Saumon}, {Freedman}, {Sudarsky}, \& {Sharp}}]{burrows97}
{Burrows}, A., {Marley}, M., {Hubbard}, W.~B., {et~al.} 1997, \bibinfo{title}{{A Nongray Theory of Extrasolar Giant Planets and Brown Dwarfs},} \apj, 491, 856, \dodoi{10.1086/305002}

\bibitem[{C. {Buzard} {et~al.}(2022){Buzard}, {Casewell}, {Lothringer}, \& {Blake}}]{Buzard22}
{Buzard}, C., {Casewell}, S.~L., {Lothringer}, J.~D., \& {Blake}, G.~A. 2022, \bibinfo{title}{{Near-infrared Spectra of the Inflated Post-common Envelope Brown Dwarf NLTT 5306 B},} \aj, 163, 262, \dodoi{10.3847/1538-3881/ac6508}

\bibitem[{T.~W. {Carmichael} {et~al.}(2022){Carmichael}, {Irwin}, {Murgas}, {Pall{\'e}}, {Stassun}, {Bartnik}, {Collins}, {de Leon}, {Esparza-Borges}, {Fedewa}, {Fong}, {Fukui}, {Jenkins}, {Kagetani}, {Latham}, {Lund}, {Mann}, {Moldovan}, {Morgan}, {Narita}, {Painter}, {Parviainen}, {Quintana}, {Ricker}, {Schulte}, {Schwarz}, {Seager}, {Sokolovsky}, {Twicken}, \& {Winn}}]{Carmichael22}
{Carmichael}, T.~W., {Irwin}, J.~M., {Murgas}, F., {et~al.} 2022, \bibinfo{title}{{TOI-2119: a transiting brown dwarf orbiting an active M-dwarf from NASA's TESS mission},} \mnras, 514, 4944, \dodoi{10.1093/mnras/stac1666}

\bibitem[{S.~L. {Casewell} {et~al.}(2020{\natexlab{a}}){Casewell}, {Debes}, {Braker}, {Cushing}, {Mace}, {Marley}, \& {Kirkpatrick}}]{Casewell20}
{Casewell}, S.~L., {Debes}, J., {Braker}, I.~P., {et~al.} 2020{\natexlab{a}}, \bibinfo{title}{{NLTT5306B: an inflated, weakly irradiated brown dwarf},} \mnras, 499, 5318, \dodoi{10.1093/mnras/staa3184}

\bibitem[{S.~L. {Casewell} {et~al.}(2018){Casewell}, {Braker}, {Parsons}, {Hermes}, {Burleigh}, {Belardi}, {Chaushev}, {Finch}, {Roy}, {Littlefair}, {Goad}, \& {Dennihy}}]{Casewell18}
{Casewell}, S.~L., {Braker}, I.~P., {Parsons}, S.~G., {et~al.} 2018, \bibinfo{title}{{The first sub-70 min non-interacting WD-BD system: EPIC212235321},} \mnras, 476, 1405, \dodoi{10.1093/mnras/sty245}

\bibitem[{S.~L. {Casewell} {et~al.}(2020{\natexlab{b}}){Casewell}, {Belardi}, {Parsons}, {Littlefair}, {Braker}, {Hermes}, {Debes}, {Vanderbosch}, {Burleigh}, {G{\"a}nsicke}, {Dhillon}, {Marsh}, {Winget}, \& {Winget}}]{Casewell20b}
{Casewell}, S.~L., {Belardi}, C., {Parsons}, S.~G., {et~al.} 2020{\natexlab{b}}, \bibinfo{title}{{WD1032 + 011, an inflated brown dwarf in an old eclipsing binary with a white dwarf},} \mnras, 497, 3571, \dodoi{10.1093/mnras/staa1608}

\bibitem[{S.~L. {Casewell} {et~al.}(2024){Casewell}, {Burleigh}, {Napiwotzki}, {Zorotovic}, {Bergeron}, {French}, {Hermes}, {Faedi}, \& {Lawrie}}]{Casewell2024}
{Casewell}, S.~L., {Burleigh}, M.~R., {Napiwotzki}, R., {et~al.} 2024, \bibinfo{title}{{The evolutionary history of GD 1400AB, a white dwarf-brown dwarf binary},} \mnras, 535, 753, \dodoi{10.1093/mnras/stae2301}

\bibitem[{G. {Chabrier} {et~al.}(2005){Chabrier}, {Baraffe}, {Allard}, \& {Hauschildt}}]{chabrier2005}
{Chabrier}, G., {Baraffe}, I., {Allard}, F., \& {Hauschildt}, P.~H. 2005, \bibinfo{title}{{Review on low-mass stars and brown dwarfs},} arXiv e-prints, astro, \dodoi{10.48550/arXiv.astro-ph/0509798}

\bibitem[{G. {Chabrier} {et~al.}(2023){Chabrier}, {Baraffe}, {Phillips}, \& {Debras}}]{chabrier23}
{Chabrier}, G., {Baraffe}, I., {Phillips}, M., \& {Debras}, F. 2023, \bibinfo{title}{{Impact of a new H/He equation of state on the evolution of massive brown dwarfs. New determination of the hydrogen burning limit},} \aap, 671, A119, \dodoi{10.1051/0004-6361/202243832}

\bibitem[{G. {Chabrier} {et~al.}(2019){Chabrier}, {Mazevet}, \& {Soubiran}}]{cms19}
{Chabrier}, G., {Mazevet}, S., \& {Soubiran}, F. 2019, \bibinfo{title}{{A New Equation of State for Dense Hydrogen-Helium Mixtures},} \apj, 872, 51, \dodoi{10.3847/1538-4357/aaf99f}

\bibitem[{Y. {Chachan} {et~al.}(2025){Chachan}, {Fortney}, {Ohno}, {Thorngren}, \& {Murray-Clay}}]{Chachan2025}
{Chachan}, Y., {Fortney}, J.~J., {Ohno}, K., {Thorngren}, D., \& {Murray-Clay}, R. 2025, \bibinfo{title}{{Revising the Giant Planet Mass─Metallicity Relation: Deciphering the Formation Sequence of Giant Planets},} \apj, 994, 43, \dodoi{10.3847/1538-4357/ae0cbf}

\bibitem[{C.~E. {Davis} {et~al.}(2025){Davis}, {Fortney}, {Iyer}, {Mukherjee}, {Morley}, {Marley}, {Line}, \& {Muirhead}}]{evan2025}
{Davis}, C.~E., {Fortney}, J.~J., {Iyer}, A., {et~al.} 2025, \bibinfo{title}{{The Sonora Substellar Atmosphere Models VI. Red Diamondback: Extending Diamondback with SPHINX for Brown Dwarf Early Evolution},} arXiv e-prints, arXiv:2510.08694, \dodoi{10.48550/arXiv.2510.08694}

\bibitem[{B. {Drummond} {et~al.}(2016){Drummond}, {Tremblin}, {Baraffe}, {Amundsen}, {Mayne}, {Venot}, \& {Goyal}}]{drummond16}
{Drummond}, B., {Tremblin}, P., {Baraffe}, I., {et~al.} 2016, \bibinfo{title}{{The effects of consistent chemical kinetics calculations on the pressure-temperature profiles and emission spectra of hot Jupiters},} \aap, 594, A69, \dodoi{10.1051/0004-6361/201628799}

\bibitem[{T.~J. {Dupuy} \& A.~L. {Kraus}(2013){Dupuy} \& {Kraus}}]{dupuy13}
{Dupuy}, T.~J., \& {Kraus}, A.~L. 2013, \bibinfo{title}{{Distances, Luminosities, and Temperatures of the Coldest Known Substellar Objects},} Science, 341, 1492, \dodoi{10.1126/science.1241917}

\bibitem[{J.~J. {Fortney} {et~al.}(2008){Fortney}, {Lodders}, {Marley}, \& {Freedman}}]{fortney08}
{Fortney}, J.~J., {Lodders}, K., {Marley}, M.~S., \& {Freedman}, R.~S. 2008, \bibinfo{title}{{A Unified Theory for the Atmospheres of the Hot and Very Hot Jupiters: Two Classes of Irradiated Atmospheres},} \apj, 678, 1419, \dodoi{10.1086/528370}

\bibitem[{J.~J. {Fortney} {et~al.}(2007){Fortney}, {Marley}, \& {Barnes}}]{fortney2007planetary}
{Fortney}, J.~J., {Marley}, M.~S., \& {Barnes}, J.~W. 2007, \bibinfo{title}{{Planetary Radii across Five Orders of Magnitude in Mass and Stellar Insolation: Application to Transits},} \apj, 659, 1661, \dodoi{10.1086/512120}

\bibitem[{J.~J. {Fortney} \& N. {Nettelmann}(2010){Fortney} \& {Nettelmann}}]{fortney10}
{Fortney}, J.~J., \& {Nettelmann}, N. 2010, \bibinfo{title}{{The Interior Structure, Composition, and Evolution of Giant Planets},} \ssr, 152, 423, \dodoi{10.1007/s11214-009-9582-x}

\bibitem[{J.~J. {Fortney} {et~al.}(2020){Fortney}, {Visscher}, {Marley}, {Hood}, {Line}, {Thorngren}, {Freedman}, \& {Lupu}}]{fortney20}
{Fortney}, J.~J., {Visscher}, C., {Marley}, M.~S., {et~al.} 2020, \bibinfo{title}{{Beyond Equilibrium Temperature: How the Atmosphere/Interior Connection Affects the Onset of Methane, Ammonia, and Clouds in Warm Transiting Giant Planets},} \aj, 160, 288, \dodoi{10.3847/1538-3881/abc5bd}

\bibitem[{G. {Fu} {et~al.}(2025){Fu}, {Mukherjee}, {Stevenson}, {Sing}, {Ashtari}, {Mayne}, {Lothringer}, {Zamyatina}, {Schmidt}, {Gasc{\'o}n}, {Allen}, {Bennett}, \& {L{\'o}pez-Morales}}]{fu25limb}
{Fu}, G., {Mukherjee}, S., {Stevenson}, K.~B., {et~al.} 2025, \bibinfo{title}{{Overcast mornings and clear evenings in hot Jupiter exoplanet atmospheres},} arXiv e-prints, arXiv:2507.15854, \dodoi{10.48550/arXiv.2507.15854}

\bibitem[{P. {Gao} {et~al.}(2017){Gao}, {Marley}, {Zahnle}, {Robinson}, \& {Lewis}}]{gao2017sulfur}
{Gao}, P., {Marley}, M.~S., {Zahnle}, K., {Robinson}, T.~D., \& {Lewis}, N.~K. 2017, \bibinfo{title}{{Sulfur Hazes in Giant Exoplanet Atmospheres: Impacts on Reflected Light Spectra},} \aj, 153, 139, \dodoi{10.3847/1538-3881/aa5fab}

\bibitem[{D. {Grether} \& C.~H. {Lineweaver}(2006){Grether} \& {Lineweaver}}]{grether06}
{Grether}, D., \& {Lineweaver}, C.~H. 2006, \bibinfo{title}{{How Dry is the Brown Dwarf Desert? Quantifying the Relative Number of Planets, Brown Dwarfs, and Stellar Companions around Nearby Sun-like Stars},} \apj, 640, 1051, \dodoi{10.1086/500161}

\bibitem[{C. {Helling} {et~al.}(2008){Helling}, {Ackerman}, {Allard}, {Dehn}, {Hauschildt}, {Homeier}, {Lodders}, {Marley}, {Rietmeijer}, {Tsuji}, \& {Woitke}}]{helling08}
{Helling}, C., {Ackerman}, A., {Allard}, F., {et~al.} 2008, \bibinfo{title}{{A comparison of chemistry and dust cloud formation in ultracool dwarf model atmospheres},} \mnras, 391, 1854, \dodoi{10.1111/j.1365-2966.2008.13991.x}

\bibitem[{B.~A. {Henderson} {et~al.}(2024{\natexlab{a}}){Henderson}, {Casewell}, {Goad}, {Acton}, {G{\"u}nther}, {Nielsen}, {Burleigh}, {Belardi}, {Tilbrook}, {Turner}, {Howell}, {Clark}, {Littlefield}, {Barkaoui}, {Alves}, {Anderson}, {Bayliss}, {Bouchy}, {Bryant}, {Dransfield}, {Ducrot}, {Eigm{\"u}ller}, {Gill}, {Gillen}, {Gillon}, {Hawthorn}, {Hooton}, {Jackman}, {Jehin}, {Jenkins}, {Kendall}, {Lendl}, {McCormac}, {Moyano}, {Pedersen}, {Pozuelos}, {Ramsay}, {Sefako}, {Timmermans}, {Triaud}, {Udry}, {Vines}, {Watson}, {West}, {Wheatley}, \& {Z{\'u}{\~n}iga-Fern{\'a}ndez}}]{Henderson24}
{Henderson}, B.~A., {Casewell}, S.~L., {Goad}, M.~R., {et~al.} 2024{\natexlab{a}}, \bibinfo{title}{{NGTS-28Ab: a short period transiting brown dwarf},} \mnras, 530, 318, \dodoi{10.1093/mnras/stae508}

\bibitem[{B.~A. {Henderson} {et~al.}(2024{\natexlab{b}}){Henderson}, {Casewell}, {Jord{\'a}n}, {Brahm}, {Henning}, {Gill}, {Mayorga}, {Ziegler}, {Stassun}, {Goad}, {Acton}, {Alves}, {Anderson}, {Apergis}, {Armstrong}, {Bayliss}, {Burleigh}, {Dragomir}, {Gillen}, {G{\"u}nther}, {Hedges}, {Hesse}, {Hobson}, {Jenkins}, {Jenkins}, {Kendall}, {Lendl}, {Lund}, {McCormac}, {Moyano}, {Osborn}, {Pinto}, {Ramsay}, {Rapetti}, {Saha}, {Seager}, {Trifonov}, {Udry}, {Vines}, {West}, {Wheatley}, {Winn}, \& {Zivave}}]{henderson2024}
{Henderson}, B.~A., {Casewell}, S.~L., {Jord{\'a}n}, A., {et~al.} 2024{\natexlab{b}}, \bibinfo{title}{{TOI-2490b - the most eccentric brown dwarf transiting in the brown dwarf desert},} \mnras, 533, 2823, \dodoi{10.1093/mnras/stae1940}

\bibitem[{K.~K.~W. {Hoch} {et~al.}(2025){Hoch}, {Rowland}, {Petrus}, {Nasedkin}, {Ingebretsen}, {Kammerer}, {Perrin}, {D'Orazi}, {Balmer}, {Barman}, {Bonnefoy}, {Chauvin}, {Chen}, {De Rosa}, {Girard}, {Gonzales}, {Kenworthy}, {Konopacky}, {Macintosh}, {Moran}, {Morley}, {Palma-Bifani}, {Pueyo}, {Ren}, {Rickman}, {Ruffio}, {Theissen}, {Ward-Duong}, \& {Zhang}}]{hoch25}
{Hoch}, K.~K.~W., {Rowland}, M., {Petrus}, S., {et~al.} 2025, \bibinfo{title}{{Silicate clouds and a circumplanetary disk in the YSES-1 exoplanet system},} \nat, 643, 938, \dodoi{10.1038/s41586-025-09174-w}

\bibitem[{I. {Hubeny} {et~al.}(2003){Hubeny}, {Burrows}, \& {Sudarsky}}]{hubeny03}
{Hubeny}, I., {Burrows}, A., \& {Sudarsky}, D. 2003, \bibinfo{title}{{A Possible Bifurcation in Atmospheres of Strongly Irradiated Stars and Planets},} \apj, 594, 1011, \dodoi{10.1086/377080}

\bibitem[{J.~D. Hunter(2007)Hunter}]{Hunter:2007}
Hunter, J.~D. 2007, \bibinfo{title}{Matplotlib: A 2D graphics environment,} Computing in Science \& Engineering, 9, 90, \dodoi{10.1109/MCSE.2007.55}

\bibitem[{J. {Inglis} {et~al.}(2024){Inglis}, {Batalha}, {Lewis}, {Kataria}, {Knutson}, {Kilpatrick}, {Gagnebin}, {Mukherjee}, {Pettyjohn}, {Crossfield}, {Foote}, {Grant}, {Henry}, {Lally}, {McKemmish}, {Sing}, {Wakeford}, {Zapata Trujillo}, \& {Zellem}}]{inglis24}
{Inglis}, J., {Batalha}, N.~E., {Lewis}, N.~K., {et~al.} 2024, \bibinfo{title}{{Quartz Clouds in the Dayside Atmosphere of the Quintessential Hot Jupiter HD 189733 b},} \apjl, 973, L41, \dodoi{10.3847/2041-8213/ad725e}

\bibitem[{Y. {Kawashima} \& M. {Ikoma}(2018){Kawashima} \& {Ikoma}}]{kawashima18}
{Kawashima}, Y., \& {Ikoma}, M. 2018, \bibinfo{title}{{Theoretical Transmission Spectra of Exoplanet Atmospheres with Hydrocarbon Haze: Effect of Creation, Growth, and Settling of Haze Particles. I. Model Description and First Results},} \apj, 853, 7, \dodoi{10.3847/1538-4357/aaa0c5}

\bibitem[{T. {Kluyver} {et~al.}(2016){Kluyver}, {Ragan-Kelley}, {P{\'e}rez}, {Granger}, {Bussonnier}, {Frederic}, {Kelley}, {Hamrick}, {Grout}, {Corlay}, {Ivanov}, {Avila}, {Abdalla}, {Willing}, \& {Jupyter Development Team}}]{Kluyver}
{Kluyver}, T., {Ragan-Kelley}, B., {P{\'e}rez}, F., {et~al.} 2016, \bibinfo{title}{{Jupyter Notebooks{\textemdash}a publishing format for reproducible computational workflows},} in IOS Press, 87--90, \dodoi{10.3233/978-1-61499-649-1-87}

\bibitem[{T.~D. {Komacek} \& A.~N. {Youdin}(2017){Komacek} \& {Youdin}}]{komacek17}
{Komacek}, T.~D., \& {Youdin}, A.~N. 2017, \bibinfo{title}{{Structure and Evolution of Internally Heated Hot Jupiters},} \apj, 844, 94, \dodoi{10.3847/1538-4357/aa7b75}

\bibitem[{S.~S. {Kumar}(1963){Kumar}}]{kumar63}
{Kumar}, S.~S. 1963, \bibinfo{title}{{The Structure of Stars of Very Low Mass.},} \apj, 137, 1121, \dodoi{10.1086/147589}

\bibitem[{B.~W.~P. {Lew} {et~al.}(2022){Lew}, {Apai}, {Zhou}, {Marley}, {Mayorga}, {Tan}, {Parmentier}, {Casewell}, \& {Xu (许偲艺)}}]{Lew22}
{Lew}, B. W.~P., {Apai}, D., {Zhou}, Y., {et~al.} 2022, \bibinfo{title}{{Mapping the Pressure-dependent Day-Night Temperature Contrast of a Strongly Irradiated Atmosphere with HST Spectroscopic Phase Curve},} \aj, 163, 8, \dodoi{10.3847/1538-3881/ac3001}

\bibitem[{K. {Lodders} {et~al.}(2009){Lodders}, {Palme}, \& {Gail}}]{lodders09}
{Lodders}, K., {Palme}, H., \& {Gail}, H.~P. 2009, \bibinfo{title}{{Abundances of the Elements in the Solar System},} Landolt B\&ouml;rnstein, 4B, 712, \dodoi{10.1007/978-3-540-88055-4_34}

\bibitem[{E.~D. {Lopez} \& J.~J. {Fortney}(2014){Lopez} \& {Fortney}}]{lopez14}
{Lopez}, E.~D., \& {Fortney}, J.~J. 2014, \bibinfo{title}{{Understanding the Mass-Radius Relation for Sub-neptunes: Radius as a Proxy for Composition},} \apj, 792, 1, \dodoi{10.1088/0004-637X/792/1/1}

\bibitem[{J.~D. {Lothringer} \& S.~L. {Casewell}(2020){Lothringer} \& {Casewell}}]{lothringer20}
{Lothringer}, J.~D., \& {Casewell}, S.~L. 2020, \bibinfo{title}{{Atmosphere Models of Brown Dwarfs Irradiated by White Dwarfs: Analogs for Hot and Ultrahot Jupiters},} \apj, 905, 163, \dodoi{10.3847/1538-4357/abc5bc}

\bibitem[{M.~S. {Marley} {et~al.}(2021){Marley}, {Saumon}, {Visscher}, {Lupu}, {Freedman}, {Morley}, {Fortney}, {Seay}, {Smith}, {Teal}, \& {Wang}}]{marley21}
{Marley}, M.~S., {Saumon}, D., {Visscher}, C., {et~al.} 2021, \bibinfo{title}{{The Sonora Brown Dwarf Atmosphere and Evolution Models. I. Model Description and Application to Cloudless Atmospheres in Rainout Chemical Equilibrium},} \apj, 920, 85, \dodoi{10.3847/1538-4357/ac141d}

\bibitem[{K. {Menou}(2012){Menou}}]{Menou12}
{Menou}, K. 2012, \bibinfo{title}{{Magnetic Scaling Laws for the Atmospheres of Hot Giant Exoplanets},} \apj, 745, 138, \dodoi{10.1088/0004-637X/745/2/138}

\bibitem[{Y. {Miguel} \& A. {Vazan}(2023){Miguel} \& {Vazan}}]{miguel23}
{Miguel}, Y., \& {Vazan}, A. 2023, \bibinfo{title}{{Interior and Evolution of the Giant Planets},} Remote Sensing, 15, 681, \dodoi{10.3390/rs15030681}

\bibitem[{B.~E. {Miles} {et~al.}(2022){Miles}, {Biller}, {Patapis}, {Worthen}, {Rickman}, {Hoch}, {Skemer}, {Perrin}, {Whiteford}, {Chen}, {Mukherjee}, {Morley}, {Moran}, {Bonnefoy}, {Petrus}, {Carter}, {Choquet}, {Hinkley}, {Ward-Duong}, {Leisenring}, {Millar-Blanchaer}, {Pueyo}, {Ray}, {Stapelfeldt}, {Stone}, {Wang}, {Absil}, {Balmer}, {Boccaletti}, {Bonavita}, {Booth}, {Bowler}, {Chauvin}, {Christiaens}, {Currie}, {Danielski}, {Fortney}, {Girard}, {Greenbaum}, {Henning}, {Hines}, {Janson}, {Kalas}, {Kammerer}, {Kenworthy}, {Kervella}, {Lagage}, {Lew}, {Liu}, {Macintosh}, {Marino}, {Marley}, {Marois}, {Matthews}, {Matthews}, {Mawet}, {McElwain}, {Metchev}, {Meyer}, {Molliere}, {Pantin}, {Rebollido}, {Ren}, {Vasist}, {Wyatt}, {Zhou}, {Briesemeister}, {Bryan}, {Calissendorff}, {Catalloube}, {Cugno}, {De Furio}, {Dupuy}, {Factor}, {Faherty}, {Fitzgerald}, {Franson}, {Gonzales}, {Hood}, {Howe}, {Kraus}, {Kuzuhara}, {Lawson}, {Lazzoni}, {Liu}, {Llop-Sayson}, {Lloyd}, {Martinez}, {Mazoyer}, {Quanz}, {Adams
  Redai}, {Samland}, {Schlieder}, {Tamura}, {Tan}, {Uyama}, {Vigan}, {Vos}, {Wagner}, {Wolff}, {Ygouf}, {Zhang}, \& {Zhang}}]{miles22}
{Miles}, B.~E., {Biller}, B.~A., {Patapis}, P., {et~al.} 2022, \bibinfo{title}{{The JWST Early Release Science Program for Direct Observations of Exoplanetary Systems II: A 1 to 20 Micron Spectrum of the Planetary-Mass Companion VHS 1256-1257 b},} arXiv e-prints, arXiv:2209.00620, \dodoi{10.48550/arXiv.2209.00620}

\bibitem[{P. {Molli{\`e}re} \& C. {Mordasini}(2012){Molli{\`e}re} \& {Mordasini}}]{molliere12}
{Molli{\`e}re}, P., \& {Mordasini}, C. 2012, \bibinfo{title}{{Deuterium burning in objects forming via the core accretion scenario. Brown dwarfs or planets?},} \aap, 547, A105, \dodoi{10.1051/0004-6361/201219844}

\bibitem[{P. {Molli{\`e}re} {et~al.}(2015){Molli{\`e}re}, {van Boekel}, {Dullemond}, {Henning}, \& {Mordasini}}]{molliere15}
{Molli{\`e}re}, P., {van Boekel}, R., {Dullemond}, C., {Henning}, T., \& {Mordasini}, C. 2015, \bibinfo{title}{{Model Atmospheres of Irradiated Exoplanets: The Influence of Stellar Parameters, Metallicity, and the C/O Ratio},} \apj, 813, 47, \dodoi{10.1088/0004-637X/813/1/47}

\bibitem[{C.~V. {Morley} {et~al.}(2012){Morley}, {Fortney}, {Marley}, {Visscher}, {Saumon}, \& {Leggett}}]{morley2012neglected}
{Morley}, C.~V., {Fortney}, J.~J., {Marley}, M.~S., {et~al.} 2012, \bibinfo{title}{{Neglected Clouds in T and Y Dwarf Atmospheres},} \apj, 756, 172, \dodoi{10.1088/0004-637X/756/2/172}

\bibitem[{C.~V. {Morley} {et~al.}(2014){Morley}, {Marley}, {Fortney}, {Lupu}, {Saumon}, {Greene}, \& {Lodders}}]{morley14}
{Morley}, C.~V., {Marley}, M.~S., {Fortney}, J.~J., {et~al.} 2014, \bibinfo{title}{{Water Clouds in Y Dwarfs and Exoplanets},} \apj, 787, 78, \dodoi{10.1088/0004-637X/787/1/78}

\bibitem[{C.~V. {Morley} {et~al.}(2024){Morley}, {Mukherjee}, {Marley}, {Fortney}, {Visscher}, {Lupu}, {Gharib-Nezhad}, {Thorngren}, {Freedman}, \& {Batalha 7}}]{diamondback}
{Morley}, C.~V., {Mukherjee}, S., {Marley}, M.~S., {et~al.} 2024, \bibinfo{title}{{The Sonora Substellar Atmosphere Models. III. Diamondback: Atmospheric Properties, Spectra, and Evolution for Warm Cloudy Substellar Objects},} arXiv e-prints, arXiv:2402.00758, \dodoi{10.48550/arXiv.2402.00758}

\bibitem[{J.~I. {Moses} {et~al.}(2011){Moses}, {Visscher}, {Fortney}, {Showman}, {Lewis}, {Griffith}, {Klippenstein}, {Shabram}, {Friedson}, {Marley}, \& {Freedman}}]{moses11}
{Moses}, J.~I., {Visscher}, C., {Fortney}, J.~J., {et~al.} 2011, \bibinfo{title}{{Disequilibrium Carbon, Oxygen, and Nitrogen Chemistry in the Atmospheres of HD 189733b and HD 209458b},} \apj, 737, 15, \dodoi{10.1088/0004-637X/737/1/15}

\bibitem[{S. {Mukherjee} {et~al.}(2023){Mukherjee}, {Batalha}, {Fortney}, \& {Marley}}]{Mukherjee22}
{Mukherjee}, S., {Batalha}, N.~E., {Fortney}, J.~J., \& {Marley}, M.~S. 2023, \bibinfo{title}{{PICASO 3.0: A One-dimensional Climate Model for Giant Planets and Brown Dwarfs},} \apj, 942, 71, \dodoi{10.3847/1538-4357/ac9f48}

\bibitem[{S. {Mukherjee} {et~al.}(2022){Mukherjee}, {Fortney}, {Batalha}, {Karalidi}, {Marley}, {Visscher}, {Miles}, \& {Skemer}}]{Mukherjee22a}
{Mukherjee}, S., {Fortney}, J.~J., {Batalha}, N.~E., {et~al.} 2022, \bibinfo{title}{{Probing the Extent of Vertical Mixing in Brown Dwarf Atmospheres with Disequilibrium Chemistry},} \apj, 938, 107, \dodoi{10.3847/1538-4357/ac8dfb}

\bibitem[{S. {Mukherjee} {et~al.}(2025{\natexlab{a}}){Mukherjee}, {Fortney}, {Wogan}, {Sing}, \& {Ohno}}]{Mukherjee25a}
{Mukherjee}, S., {Fortney}, J.~J., {Wogan}, N.~F., {Sing}, D.~K., \& {Ohno}, K. 2025{\natexlab{a}}, \bibinfo{title}{{Effects of Planetary Parameters on Disequilibrium Chemistry in Irradiated Planetary Atmospheres: From Gas Giants to Sub-Neptunes},} \apj, 985, 209, \dodoi{10.3847/1538-4357/adc7b3}

\bibitem[{S. {Mukherjee} {et~al.}(2024){Mukherjee}, {Fortney}, {Morley}, {Batalha}, {Marley}, {Karalidi}, {Visscher}, {Lupu}, {Freedman}, \& {Gharib-Nezhad}}]{Mukherjee24}
{Mukherjee}, S., {Fortney}, J.~J., {Morley}, C.~V., {et~al.} 2024, \bibinfo{title}{{The Sonora Substellar Atmosphere Models. IV. Elf Owl: Atmospheric Mixing and Chemical Disequilibrium with Varying Metallicity and C/O Ratios},} \apj, 963, 73, \dodoi{10.3847/1538-4357/ad18c2}

\bibitem[{S. {Mukherjee} {et~al.}(2025{\natexlab{b}}){Mukherjee}, {Sing}, {Fu}, {Stevenson}, {Schmidt}, {Baskett}, {McCreery}, {Allen}, {Bennett}, {Christie}, {Gasc{\'o}n}, {Goyal}, {H{\'e}brard}, {Lothringer}, {L{\'o}pez-Morales}, {Lustig-Yaeger}, {May}, {Mayorga}, {Mayne}, {Ramos Rosado}, {Reggiani}, {Rustamkulov}, {Schlaufman}, {Sotzen}, {Thorngren}, {Wang}, \& {Zamyatina}}]{mukherjee25W94Ab}
{Mukherjee}, S., {Sing}, D.~K., {Fu}, G., {et~al.} 2025{\natexlab{b}}, \bibinfo{title}{{Cloudy mornings and clear evenings on a giant extrasolar world},} arXiv e-prints, arXiv:2505.10910, \dodoi{10.48550/arXiv.2505.10910}

\bibitem[{K. {Ohno} \& Y. {Kawashima}(2020){Ohno} \& {Kawashima}}]{ohno20}
{Ohno}, K., \& {Kawashima}, Y. 2020, \bibinfo{title}{{Super-Rayleigh Slopes in Transmission Spectra of Exoplanets Generated by Photochemical Haze},} \apjl, 895, L47, \dodoi{10.3847/2041-8213/ab93d7}

\bibitem[{T. pandas~development team(2020)pandas~development team}]{reback2020pandas}
pandas~development team, T. 2020, pandas-dev/pandas: Pandas, latest Zenodo, \dodoi{10.5281/zenodo.3509134}

\bibitem[{M.~W. {Phillips} {et~al.}(2020){Phillips}, {Tremblin}, {Baraffe}, {Chabrier}, {Allard}, {Spiegelman}, {Goyal}, {Drummond}, \& {H{\'e}brard}}]{Philips20}
{Phillips}, M.~W., {Tremblin}, P., {Baraffe}, I., {et~al.} 2020, \bibinfo{title}{{A new set of atmosphere and evolution models for cool T-Y brown dwarfs and giant exoplanets},} \aap, 637, A38, \dodoi{10.1051/0004-6361/201937381}

\bibitem[{T.~M. {Rogers} \& T.~D. {Komacek}(2014){Rogers} \& {Komacek}}]{rogers14}
{Rogers}, T.~M., \& {Komacek}, T.~D. 2014, \bibinfo{title}{{Magnetic Effects in Hot Jupiter Atmospheres},} \apj, 794, 132, \dodoi{10.1088/0004-637X/794/2/132}

\bibitem[{F. {Sainsbury-Martinez} {et~al.}(2021){Sainsbury-Martinez}, {Casewell}, {Lothringer}, {Phillips}, \& {Tremblin}}]{Sainsbury-Martinez2021}
{Sainsbury-Martinez}, F., {Casewell}, S.~L., {Lothringer}, J.~D., {Phillips}, M.~W., \& {Tremblin}, P. 2021, \bibinfo{title}{{Exploring deep and hot adiabats as a potential solution to the radius inflation problem in brown dwarfs. Long-timescale models of the deep atmospheres of KELT-1b, Kepler-13Ab, and SDSS1411B},} \aap, 656, A128, \dodoi{10.1051/0004-6361/202141637}

\bibitem[{P. {Sarkis} {et~al.}(2021){Sarkis}, {Mordasini}, {Henning}, {Marleau}, \& {Molli{\`e}re}}]{Sarkis21}
{Sarkis}, P., {Mordasini}, C., {Henning}, T., {Marleau}, G.~D., \& {Molli{\`e}re}, P. 2021, \bibinfo{title}{{Evidence of three mechanisms explaining the radius anomaly of hot Jupiters},} \aap, 645, A79, \dodoi{10.1051/0004-6361/202038361}

\bibitem[{D. {Saumon} \& M.~S. {Marley}(2008){Saumon} \& {Marley}}]{saumonmarley08}
{Saumon}, D., \& {Marley}, M.~S. 2008, \bibinfo{title}{{The Evolution of L and T Dwarfs in Color-Magnitude Diagrams},} \apj, 689, 1327, \dodoi{10.1086/592734}

\bibitem[{D.~K. {Sing} {et~al.}(2016){Sing}, {Fortney}, {Nikolov}, {Wakeford}, {Kataria}, {Evans}, {Aigrain}, {Ballester}, {Burrows}, {Deming}, {D{\'e}sert}, {Gibson}, {Henry}, {Huitson}, {Knutson}, {Lecavelier Des Etangs}, {Pont}, {Showman}, {Vidal-Madjar}, {Williamson}, \& {Wilson}}]{sing16}
{Sing}, D.~K., {Fortney}, J.~J., {Nikolov}, N., {et~al.} 2016, \bibinfo{title}{{A continuum from clear to cloudy hot-Jupiter exoplanets without primordial water depletion},} \nat, 529, 59, \dodoi{10.1038/nature16068}

\bibitem[{D.~K. {Sing} {et~al.}(2024){Sing}, {Rustamkulov}, {Thorngren}, {Barstow}, {Tremblin}, {Alves de Oliveira}, {Beck}, {Birkmann}, {Challener}, {Crouzet}, {Espinoza}, {Ferruit}, {Giardino}, {Gressier}, {Lee}, {Lewis}, {Maiolino}, {Manjavacas}, {Rauscher}, {Sirianni}, \& {Valenti}}]{sing24}
{Sing}, D.~K., {Rustamkulov}, Z., {Thorngren}, D.~P., {et~al.} 2024, \bibinfo{title}{{A warm Neptune's methane reveals core mass and vigorous atmospheric mixing},} arXiv e-prints, arXiv:2405.11027, \dodoi{10.48550/arXiv.2405.11027}

\bibitem[{R.~J. {Siverd} {et~al.}(2012){Siverd}, {Beatty}, {Pepper}, {Eastman}, {Collins}, {Bieryla}, {Latham}, {Buchhave}, {Jensen}, {Crepp}, {Street}, {Stassun}, {Gaudi}, {Berlind}, {Calkins}, {DePoy}, {Esquerdo}, {Fulton}, {F{\H{u}}r{\'e}sz}, {Geary}, {Gould}, {Hebb}, {Kielkopf}, {Marshall}, {Pogge}, {Stanek}, {Stefanik}, {Szentgyorgyi}, {Trueblood}, {Trueblood}, {Stutz}, \& {van Saders}}]{siverd12}
{Siverd}, R.~J., {Beatty}, T.~G., {Pepper}, J., {et~al.} 2012, \bibinfo{title}{{KELT-1b: A Strongly Irradiated, Highly Inflated, Short Period, 27 Jupiter-mass Companion Transiting a Mid-F Star},} \apj, 761, 123, \dodoi{10.1088/0004-637X/761/2/123}

\bibitem[{D.~S. {Spiegel} {et~al.}(2011){Spiegel}, {Burrows}, \& {Milsom}}]{spiegel11}
{Spiegel}, D.~S., {Burrows}, A., \& {Milsom}, J.~A. 2011, \bibinfo{title}{{The Deuterium-burning Mass Limit for Brown Dwarfs and Giant Planets},} \apj, 727, 57, \dodoi{10.1088/0004-637X/727/1/57}

\bibitem[{M.~E. {Steinrueck} {et~al.}(2023){Steinrueck}, {Koskinen}, {Lavvas}, {Parmentier}, {Zieba}, {Tan}, {Zhang}, \& {Kreidberg}}]{steinruek23}
{Steinrueck}, M.~E., {Koskinen}, T., {Lavvas}, P., {et~al.} 2023, \bibinfo{title}{{Photochemical Hazes Dramatically Alter Temperature Structure and Atmospheric Circulation in 3D Simulations of Hot Jupiters},} \apj, 951, 117, \dodoi{10.3847/1538-4357/acd4bb}

\bibitem[{S.~L. {Thompson}(1990){Thompson}}]{ANEOS}
{Thompson}, S.~L. 1990, {ANEOS---Analytic Equations of State for Shock Physics Codes, Sandia Natl. Lab. Doc. SAND89-2951}

\bibitem[{D.~P. {Thorngren}(2024){Thorngren}}]{thorngren24}
{Thorngren}, D.~P. 2024, \bibinfo{title}{{The Hot Jupiter Radius Anomaly and Stellar Connections},} arXiv e-prints, arXiv:2405.05307, \dodoi{10.48550/arXiv.2405.05307}

\bibitem[{D.~P. {Thorngren} \& J.~J. {Fortney}(2018){Thorngren} \& {Fortney}}]{thorngren2018}
{Thorngren}, D.~P., \& {Fortney}, J.~J. 2018, \bibinfo{title}{{Bayesian Analysis of Hot-Jupiter Radius Anomalies: Evidence for Ohmic Dissipation?},} \aj, 155, 214, \dodoi{10.3847/1538-3881/aaba13}

\bibitem[{D.~P. {Thorngren} {et~al.}(2021){Thorngren}, {Fortney}, {Lopez}, {Berger}, \& {Huber}}]{thorngren2021}
{Thorngren}, D.~P., {Fortney}, J.~J., {Lopez}, E.~D., {Berger}, T.~A., \& {Huber}, D. 2021, \bibinfo{title}{{Slow Cooling and Fast Reinflation for Hot Jupiters},} \apjl, 909, L16, \dodoi{10.3847/2041-8213/abe86d}

\bibitem[{D.~P. {Thorngren} {et~al.}(2016){Thorngren}, {Fortney}, {Murray-Clay}, \& {Lopez}}]{thorngren16}
{Thorngren}, D.~P., {Fortney}, J.~J., {Murray-Clay}, R.~A., \& {Lopez}, E.~D. 2016, \bibinfo{title}{{The Mass-Metallicity Relation for Giant Planets},} \apj, 831, 64, \dodoi{10.3847/0004-637X/831/1/64}

\bibitem[{C.~G. {Tinney} {et~al.}(2014){Tinney}, {Faherty}, {Kirkpatrick}, {Cushing}, {Morley}, \& {Wright}}]{tinney14}
{Tinney}, C.~G., {Faherty}, J.~K., {Kirkpatrick}, J.~D., {et~al.} 2014, \bibinfo{title}{{The Luminosities of the Coldest Brown Dwarfs},} \apj, 796, 39, \dodoi{10.1088/0004-637X/796/1/39}

\bibitem[{A.~H.~M.~J. {Triaud} {et~al.}(2017){Triaud}, {Martin}, {S{\'e}gransan}, {Smalley}, {Maxted}, {Anderson}, {Bouchy}, {Collier Cameron}, {Faedi}, {G{\'o}mez Maqueo Chew}, {Hebb}, {Hellier}, {Marmier}, {Pepe}, {Pollacco}, {Queloz}, {Udry}, \& {West}}]{triaud17}
{Triaud}, A. H.~M.~J., {Martin}, D.~V., {S{\'e}gransan}, D., {et~al.} 2017, \bibinfo{title}{{The EBLM Project. IV. Spectroscopic orbits of over 100 eclipsing M dwarfs masquerading as transiting hot Jupiters},} \aap, 608, A129, \dodoi{10.1051/0004-6361/201730993}

\bibitem[{S.-M. {Tsai} {et~al.}(2017){Tsai}, {Lyons}, {Grosheintz}, {Rimmer}, {Kitzmann}, \& {Heng}}]{tsai17}
{Tsai}, S.-M., {Lyons}, J.~R., {Grosheintz}, L., {et~al.} 2017, \bibinfo{title}{{VULCAN: An Open-source, Validated Chemical Kinetics Python Code for Exoplanetary Atmospheres},} \apjs, 228, 20, \dodoi{10.3847/1538-4365/228/2/20}

\bibitem[{S.-M. {Tsai} {et~al.}(2021){Tsai}, {Malik}, {Kitzmann}, {Lyons}, {Fateev}, {Lee}, \& {Heng}}]{tsai21}
{Tsai}, S.-M., {Malik}, M., {Kitzmann}, D., {et~al.} 2021, \bibinfo{title}{{A Comparative Study of Atmospheric Chemistry with VULCAN},} \apj, 923, 264, \dodoi{10.3847/1538-4357/ac29bc}

\bibitem[{N. {Vowell} {et~al.}(2025){Vowell}, {Rodriguez}, {Latham}, {Quinn}, {Schulte}, {Eastman}, {Bieryla}, {Barkaoui}, {Ciardi}, {Collins}, {Girardin}, {Heldridge}, {Kotten}, {Mancini}, {Murgas}, {Narita}, {Radford}, {Relles}, {Shporer}, {Soares-Furtado}, {Strakhov}, {Ziegler}, {Brice{\~n}o}, {Calkins}, {Clark}, {Collins}, {Esquerdo}, {Fajardo-Acosta}, {Fukui}, {Watkins}, {He}, {Horne}, {Jenkins}, {Mann}, {Naponiello}, {Palle}, {Schwarz}, {Seager}, {Southworth}, {Srdoc}, {Swift}, \& {Winn}}]{vowell2025}
{Vowell}, N., {Rodriguez}, J.~E., {Latham}, D.~W., {et~al.} 2025, \bibinfo{title}{{11 New Transiting Brown Dwarfs and Very Low Mass Stars from TESS},} arXiv e-prints, arXiv:2501.09795, \dodoi{10.48550/arXiv.2501.09795}

\bibitem[{L. {Welbanks} {et~al.}(2024){Welbanks}, {Bell}, {Beatty}, {Line}, {Ohno}, {Fortney}, {Schlawin}, {Greene}, {Rauscher}, {McGill}, {Murphy}, {Parmentier}, {Tang}, {Edelman}, {Mukherjee}, {Wiser}, {Lagage}, {Dyrek}, \& {Arnold}}]{welbanks24}
{Welbanks}, L., {Bell}, T.~J., {Beatty}, T.~G., {et~al.} 2024, \bibinfo{title}{{A High Internal Heat Flux and Large Core in a Warm Neptune Exoplanet},} arXiv e-prints, arXiv:2405.11018, \dodoi{10.48550/arXiv.2405.11018}

\bibitem[{K. {Zahnle} {et~al.}(2016){Zahnle}, {Marley}, {Morley}, \& {Moses}}]{zahnle16}
{Zahnle}, K., {Marley}, M.~S., {Morley}, C.~V., \& {Moses}, J.~I. 2016, \bibinfo{title}{{Photolytic Hazes in the Atmosphere of 51 Eri b},} \apj, 824, 137, \dodoi{10.3847/0004-637X/824/2/137}

\bibitem[{K.~J. {Zahnle} \& M.~S. {Marley}(2014){Zahnle} \& {Marley}}]{Zahnle14}
{Zahnle}, K.~J., \& {Marley}, M.~S. 2014, \bibinfo{title}{{Methane, Carbon Monoxide, and Ammonia in Brown Dwarfs and Self-Luminous Giant Planets},} \apj, 797, 41, \dodoi{10.1088/0004-637X/797/1/41}

\end{thebibliography}
\bibliographystyle{aasjournalv7}



\end{document}